\documentclass[11pt,prd,superscriptaddress,nofootinbib]{revtex4-1}

\usepackage{graphicx}
\usepackage{epstopdf}
\usepackage{amssymb,mathrsfs,epsfig,color, url}
\usepackage{amsmath}
\usepackage{booktabs}
\usepackage[utf8x]{inputenc}
\usepackage{colortbl}
\usepackage{adjustbox}
\usepackage{hyperref}
\usepackage{slashed}
\usepackage{fancyhdr}
\usepackage[ddmmyyyy,hhmmss]{datetime}
\usepackage{filecontents}
\usepackage{float}
\usepackage{tikz}
\usetikzlibrary{tikzmark,calc,arrows,shapes,decorations.pathreplacing}
\usepackage[justification=centering]{caption}
\usepackage{multirow}
\usepackage{cancel}

\def\sbar{\overline}
\def\stilde{\widetilde}

\numberwithin{equation}{section}

\newcommand{\uonep}{\ensuremath{U(1)^\prime}}

\newcommand{\be}{\begin{equation}}
\newcommand{\ee}{\end{equation}}
\newcommand{\bea}{\begin{eqnarray}}
\newcommand{\eea}{\end{eqnarray}}

\def\stilde{\widetilde}
\def\b{\beta}

\begin{document}

\begin{titlepage}
\vspace{0.1cm}
\begin{center}
\Large\bf\boldmath Calculating $\b$-function coefficients of  Renormalization Group Equations  \unboldmath \\[2cm]

\setlength {\baselineskip}{0.2in} {\large
  Joydeep Roy \footnote{e-mail: jdroy@itp.ac.cn}}
  \\[5mm]
{\it CAS Key Laboratory of Theoretical Physics, Institute of Theoretical Physics,\\
Chinese Academy of Sciences, Beijing 100190, China.}\\[25mm]
{\bf Abstract}\\[5mm]
\end{center}
\setlength{\baselineskip}{0.2in} 

Renormalization Group Equations (RGEs) are indispensable tool to know the behavior of physical parameters at different energy scales. They are also extremely crucial if we want to extend our known Standard Model gauge group by some extra gauge groups and the $\b$-functions are the soul of these RGEs. In literature although it is quite common to find long, final expressions of the RGEs, unfortunately it is difficult to find any pedagogical review to calculate these $\b$-function coefficients explicitly from the known formulae. Therefore in this note we shall try to explain the detail calculations of RGEs by giving some explicit examples taken from the literature. The goal and hope is to provide a hand on experience on calculating these RGEs for the young readers.

\end{titlepage}


\section{Introduction}

The problem of infinities in quantum field theories have bothered the respective physics community for a long time. Historically, the presence of infinities first appear in the calculation of electron self-energy from the perturbation theory \cite{Weinberg:1995mt}. Then it was realized that they exist when the particles involved in calculations, have both, very short wavelength (\textit{ultraviolet divergences}) and very long wavelength (\textit{infrared divergences}). Several ideas were put forward to deal with these infinities and after few decades of evolution the `renormalization' procedure has become the standardized method to tackle the problem. The idea of the renormalization is to redefine some parameters of the theory so that the infinities can be absorbed. The quantum field theories which follow this method are known as 'renormalizable' theories. Kenneth Wilson's idea of renormalization suggests that in a renormalizable quantum field theory, any parameter can be considered as an energy  scale dependent quantity and such scale dependence can be described by simple differential equations, known as \textit{renormalization group equations} (RGEs). This means, if a model which is acting under some organizing principles is specified by some parameters, at an input scale, then by \textit{running} the RGEs of masses and couplings of that model one can predict how these parameters will behave at some other energy scale. For example, let us choose an arbitrary energy scale $M$ at which a theory is said to be defined. If we want the theory to be also valid at a different scale $M'$, the corresponding changes in coupling constant ($\lambda$) and field strength ($\phi$) will keep the bare Green's function fixed, but renormalized connected \textit{n}-point Green's function $G^{n}$ will shift proportionally to the field rescaling. Therefore, if we consider $G^{n}$ as a function of $M$ and $\lambda$, the shift in the $G^{n}$ due to the change of scale would be given by \cite{Peskin:1995ev}
\be 
\label{Greenfun}
dG^n = \frac{\partial G^{(n)}}{\partial M}\delta M + \frac{\partial G^{(n)}}{\partial \lambda}\delta\lambda = n\delta\eta G^n
\ee
where $\delta M$, $\delta\lambda$ and $\delta\eta$ are the shift in scale, coupling constant and field strength respectively. Conventionally, we define two dimensionless quantities 
\be
\beta=\frac{M}{\delta M}\delta\lambda, ~~;~~ \gamma=\frac{M}{\delta M}\delta\eta.
\ee
Thus Eqn. \ref{Greenfun} becomes
\be
\label{CallSym}
\left[M\frac{\partial}{\partial M}+\beta\frac{\partial}{\partial \lambda}+n\gamma\right]G^{(n)}\left(\{x_i\},M,\lambda\right)=0
\ee
which is known as Caallan-Symanzik equation \cite{Peskin:1995ev}. Thus the evolution of coupling constant with energy scale is described by the \textit{beta function} of that coupling which is defined as
\be
\label{betafunc}
\b_{\lambda_i} \equiv  \frac{d\lambda_i}{d\rm \mu}=\sum_{n=1}^{\infty}\frac{1}{(16\pi^2)^n}\b_{\lambda_i}^n=\frac{1}{16\pi^2} b_i\lambda^3+\mathcal{O}(\lambda^5)
\ee
where $\mu$ represents the renormalization energy scale and $b_i$s are the beta function coefficients determined from loop corrections. These equations play very important role in renormalization theory and they are obtained by calculating the loop-corrections of the corresponding theory. For Standard Model (SM) or Minimal Supersymmetric Standard Model (MSSM) parameters the one-loop and two-loop or even more loop order  beta functions are known for some time. These calculations are renormalization scheme-dependent and the results are available in literature. 

Model-independent generic formulae for calculating the $\beta$-function coefficients of gauge couplings in both SUSY and non-SUSY model is given by \cite{Machacek:1983tz}
\begin{eqnarray}
\beta (g) = \mu \dfrac{d g}{d\mu}&=&-\dfrac{g^3}{(4 \pi)^2}\left\lbrace\dfrac{11}{3}C_2 (G)-\dfrac{4}{3}\kappa S_2(F)-\dfrac{1}{6}\eta S_2(S)\right\rbrace  \nonumber\\ 
&& -\dfrac{g^5}{(4 \pi)^4}\left\lbrace \dfrac{34}{3} \left[ C_2(G)\right] ^2-\kappa\left[4 C_2(F) +\dfrac{20}{3}C_2(G)\right]S_2(F)\right. \nonumber\\ 
&& \left.-\left[2C_2(S)+\dfrac{1}{3} C_2(G)\right]\eta S_2(S)\right\rbrace . 
\end{eqnarray}
\noindent
where $C_2$ is the quadratic Casimir invariants for gauge multiplets ($G$), scalars ($S$) and fermions ($F$) respectively. $S_2(F)$ and $S_2(S)$ are the Dynkin indices for fermion and scalar representations, respectively.
$\kappa=1,\frac{1}{2}$ for Dirac and Weyl fermions and $\eta=1,2$ for real and complex scalar fields respectively.

\noindent
From the $\beta$-function expression we get,
\begin{eqnarray}
b_i&=&-\dfrac{11}{3}C_2 (G_i)+\dfrac{4}{3}\kappa S_2(F_i)+\dfrac{1}{6}\eta S_2(S_i)\label{genbcoeff}\\
B_{ij}&=&-\dfrac{34}{3} \left[ C_2(G_i)\right] ^2 \delta_{ij}+\kappa\left[4 C_2(F_j) +\dfrac{20}{3}\delta_{ij} C_2(G_i)\right]S_2(F_i) \nonumber\\ 
&&+\eta \left[2C_2(S_j)+\dfrac{1}{3} \delta_{ij}  C_2(G_i)\right]S_2(S_i)\label{genBcoeff} .
\end{eqnarray}

The Dynkin index, $S(R)$, for an irreducible representation $R$, defined by
\be \label{eq:defDynkin}
\mathrm{Tr}[S^aS^a] = S(R_i)\delta^{ab}
\ee
where the trace is taken only over $R_i$. For a fundamental representation of $\rm SU(N)$ the generators are usually normalized to $S(R) = 1/2$. Quadratic Casimir invariants $C_2(R)$ and the Dynkin Indices are related by the relation
\be \label{CasDyn}
d(G)S(R) = d(R)C_2(R)
\ee
where $d(R)$ is the dimension of $R$, and $d(G)$ is the dimension of the adjoint representation of the gauge group G which is simply given by the number of generators of the group. 

The amount of research works, in both supersymmetric and non-supersymmetric scenarios, that use RGEs for their purpose is significantly heavy and it is only possible to calculate few of them from both spectrum for the pedagogical purpose. Therefore in this note we first plan to calculate the beta functions SM RGEs in section \ref{SMRGEs}, starting from the generic formulae stated above. Next, in section \ref{MSSMRGEs} we calculate the same for the MSSM. We take a specific example of $\uonep$ extension of the MSSM and show the similar calculation in section \ref{UpRGEs}. Vector-like leptons are another popular extension of SM, that we shall consider in section \ref{VLRGEs}. Finally, we shall calculate the $\beta$-function coefficients of the Pati-Salam model RGEs in section \ref{PSRGEs}.


\section{$\beta$-function of SM gauge coupling RGEs (non-SUSY scenario)}
\label{SMRGEs}

The SM particle content is given by

\begin{table}[H]
\begin{center}
\begin{tabular}{|c|c|c|}
\hline
\multicolumn{2}{|c|}{Names} 
 & $SU(3)_C ,\, SU(2)_L ,\, U(1)_Y$
\\  \hline\hline
Quarks & $Q$ & $(\>{\bf 3},\>{\bf 2}\>,\>{1\over 6})$
\\
($\times 3$ families) & $\sbar u$ & 
$(\>{\bf \overline 3},\> {\bf 1},\> -{2\over 3})$
\\ & $\sbar d$ & $(\>{\bf \overline 3},\> {\bf 1},\> {1\over 3})$ \\  \hline
Leptons & $L$ & $(\>{\bf 1},\>{\bf 2}\>,\>-{1\over 2})$
\\($\times 3$ families) & $\sbar e$ & $(\>{\bf 1},\> {\bf 1},\>1)$ \\  \hline
Higgs & $H$ & $(\>{\bf 1},\>{\bf 2}\>,\> 1)$
\\  \hline
\end{tabular}
\caption{SM particle content.\label{tab:SM}}
\vspace{-0.6cm}
\end{center}
\end{table}


First, we summarize the renormalization group equations in the SM. The two-loop equations which are generally accepted and widely used for the gauge couplings are given by \cite{Martin:1993zk}
\begin{eqnarray}
(4\pi)^2\frac{d}{dt}~ g_i=g_i^3b_i &+&\frac{g_i^3}{(4\pi)^2}
\left[ \sum_{j=1}^3B_{ij}g_j^2-\sum_{\alpha=u,d,e}d_i^\alpha
{\rm Tr}\left( h^{\alpha \dagger}h^{\alpha}\right) \right] ~,~\,
\label{SMgauge}
\end{eqnarray}
where $t=\ln  \mu$ and $ \mu$ is the renormalization scale.
$g_1$, $g_2$ and $g_3$ are the gauge couplings
for $U(1)_Y$, $SU(2)_L$ and $SU(3)_C$, respectively,
where we use the $SU(5)$ normalization $g_1^2 \equiv (5/3)g_Y^{ 2}$ and $d_i^\alpha$ are the Yukawa couplings contributions.

For the Standard Model (SM), the 1-loop $\b$-function coefficients $b_i$s are very important and are
determined by the gauge group and the matter multiplets to which the gauge bosons couple. They can be found in the literature as \cite{Machacek:1983tz}
\begin{eqnarray}
\label{bfunccoeffSM}
&&(b_1, b_2, b_3)_{\mathrm{SM}}=\bigg(U(1)_Y, SU(2)_L, SU(3)_C\bigg)_{\mathrm{SM}} =  \left(\frac{41}{10},-\frac{19}{6},-7\right) 
\end{eqnarray}
\noindent
The 2-loop gauge $\b$-function coefficients are given by \cite{Machacek:1983tz, Martin:1993zk}
\be
\label{BSMcoeff}
\big(B_{ij}\big)_{\rm SM}=\begin{pmatrix}
\frac{199}{50}&
~~\frac{27}{10}& ~~\frac{44}{5}\cr \frac{9}{10} & ~~\frac{35}{6}& ~~12 \cr
\frac{11}{10}& ~~\frac{9}{2}& ~~-26
\end{pmatrix}, 
\ee
and Yukawa $\b$-function coefficients are given by
\bea
\label{YukSMcoeff}
&&\big(d^u\big)_{\rm{SM}}=\left(\frac{17}{10},\frac{3}{2},2\right) ~,~
\big(d^d\big)_{\rm{SM}}=\left(\frac{1}{2},\frac{3}{2},2\right) ~,~
\big(d^e\big)_{\rm{SM}}=\left(\frac{3}{2},\frac{1}{2},0\right) ~.~\,
\eea
Now we shall see how the numbers in equations~\ref{bfunccoeffSM} and \ref{BSMcoeff} are obtained.


\subsection{1-loop gauge $\b$-function coefficients}

To calculate these coefficients two very important parameters are quadratic Casimir invariants and Dynkin indices. Let us specify them first. The quadratic Casimir invariant for the groups are given by
\bea
&&
C_2(G) =
\Biggl \{ \begin{array}{ll}
0 & {\rm for}\>\, U(1),
\\
N & {~\rm for}\>\, SU(N).
\end{array}
\label{CasGrp}
\eea

\noindent
The quadratic Casimir invariants of SM multiplets in different representations are given by the relation $C_2(R)=\frac{N^2-1}{2N}$. Thus,
\bea
&&
C^3_R(i) =
\Biggl \{ \begin{array}{ll}
\frac{4}{3} & {~\rm for}\>\,\Phi_i = Q, \sbar u, \sbar d,
\\
0 & {~\rm for}\>\,\Phi_i = L, \sbar e, H
\end{array}
\label{defC1}
\\
&&
C^2_R(i) =
\Biggl \{ \begin{array}{ll}
\frac{3}{4} & {~\rm for}\>\,\Phi_i = Q, L, H\\
0 & {~\rm for}\>\,\Phi_i = \sbar u, \sbar d, \sbar e
,\end{array}
\label{defC2}
\\
&&
C^1_R(i) = \>
3 Y_i^2/5 \>\>\>{\rm for~each}\>\,\Phi_i\>\,{\rm
with~weak~hypercharge}\>\, Y_i.
\label{defC3}
\eea

\noindent
The Dynkin indices for the groups are 

\bea
&&
S_2(R)\equiv S(R) =
\Biggl \{ \begin{array}{ll}
\frac{3}{5}Y_{\Phi_i}^2 & {~\rm for}\>\, U(1),
\\
\frac{1}{2} & {\rm for}\>\, SU(N).
\end{array}
\label{DynGrp}
\eea

From equation \ref{genbcoeff}, using \ref{CasGrp} and \ref{DynGrp}, we can simplify the formula for calculating these coefficients as follows. For $\rm U(1)$, the contributions from all multiplets come through their hypercharge. Therefore,
\bea
\label{b1funcSM}
b_1 &=&0+\frac{4}{3}\cdot\frac{1}{\tikzmark{b1SM1}2}\cdot\bigg(\sum_{\mathrm{fermions}}\frac{3}{5}Y_{\mathrm{fermions}}^2\bigg)+\frac{1}{6}\cdot\tikzmark{b1SM2}2\cdot\bigg(\sum_{\mathrm{scalars}}\frac{3}{5}Y_{\mathrm{scalars}}^2\bigg)~~[\mathrm{for}~ U(1), C_2(G)=0]\nonumber\\
&=& \frac{2}{3}\cdot\sum_{\mathrm{fermions}}\frac{3}{5}Y_{\mathrm{fermions}}^2+\frac{1}{3}\cdot\sum_{\mathrm{scalars}}\frac{3}{5}Y_{\mathrm{scalars}}^2 \nonumber\\
&=& \frac{2}{5}\bigg[\tikzmark{b1SM3}(3\cdot\tikzmark{b1SM4}2)\cdot(\frac{1}{6})^2+(3\cdot 1)\cdot(-\frac{2}{3})^2+(3\cdot 1)\cdot(\frac{1}{3})^2+\tikzmark{b1SM5}2\cdot(-\frac{1}{2})^2+\tikzmark{b1SM6}1\cdot 1^2\bigg]N_f+ \frac{1}{5}\bigg[2\cdot(-\frac{1}{2})^2\bigg]N_H \nonumber\\
\begin{tikzpicture}[remember picture,overlay]
\draw[<-] 
  ([shift={(2pt,-2pt)}]pic cs:b1SM1) |- ([shift={(-14pt,-12pt)}]pic cs:b1SM1) 
  node[anchor=east] {$\scriptstyle \text{Weyl fermion}$}; 
\draw[<-] 
  ([shift={(2pt,-2pt)}]pic cs:b1SM2) |- ([shift={(-14pt,-14pt)}]pic cs:b1SM2) 
  node[anchor=east] {$\scriptstyle \text{Complex scalar}$};   
\draw[<-] 
  ([shift={(2pt,-2pt)}]pic cs:b1SM3) |- ([shift={(-14pt,-14pt)}]pic cs:b1SM3) 
  node[anchor=east] {$\scriptstyle \text{Color}$}; 
\draw[<-] 
  ([shift={(2pt,-2pt)}]pic cs:b1SM4) |- ([shift={(16pt,-16pt)}]pic cs:b1SM4) 
  node[anchor=west] {$\scriptstyle \text{Doublet}$}; 
\draw[<-] 
  ([shift={(2pt,-2pt)}]pic cs:b1SM5) |- ([shift={(16pt,-16pt)}]pic cs:b1SM5) 
  node[anchor=west] {$\scriptstyle \text{Doublet}$}; 
\draw[<-] 
  ([shift={(2pt,-2pt)}]pic cs:b1SM6) |- ([shift={(16pt,-16pt)}]pic cs:b1SM6) 
  node[anchor=west] {$\scriptstyle \text{Singlet}$}; 
\end{tikzpicture}
&=& \frac{4}{3}N_f+\frac{1}{10}N_H 
\eea
where $N_f$ is the number of fermion generations in the SM  and $N_H$ corresponds to the number of Higgs doublet pairs. $(3\cdot 2)$, $(3\cdot 1)$ etc. in above expression represents the number of multiplets.

\noindent
In SU(2), there are 3 multiplets which transform as doublet under SU(2) representation. Those are $Q$, $L$ and the Higgs ($H$). Therefore,
\bea
\label{b2funcSM}
b_2 &=& \frac{2}{3}\cdot\frac{1}{2}\big[(\tikzmark{b2SM1}3\cdot 1)+(1\cdot 1)\big]N_f+\frac{1}{3}\cdot\frac{1}{2}(1\cdot 1)N_H-\frac{11}{3}\cdot 2
\begin{tikzpicture}[remember picture,overlay]
\draw[<-] 
  ([shift={(2pt,-2pt)}]pic cs:b2SM1) |- ([shift={(14pt,-14pt)}]pic cs:b2SM1) 
  node[anchor=west] {$\scriptstyle \text{Color}$}; 
\end{tikzpicture}
\nonumber\\
&=& \frac{4}{3}N_f+\frac{1}{6}N_H-\frac{22}{3}.
\eea
Similarly, for SU(3) only contribution comes from the SU(3) quark triplets. Thus,
\bea
\label{b3funcSM}
b_3 &=& \frac{2}{3}.\frac{1}{2}(\tikzmark{b3SM1}2+1+1)N_f-\frac{11}{3}.3
\begin{tikzpicture}[remember picture,overlay]
\draw[<-] 
  ([shift={(2pt,-2pt)}]pic cs:b3SM1) |- ([shift={(14pt,-14pt)}]pic cs:b3SM1) 
  node[anchor=west] {$\scriptstyle \text{Doublet}$}; 
\end{tikzpicture}
\nonumber\\
&=& \frac{4}{3}N_f-11,
\eea
These match with the formula for calculating these coefficients in Ref. \cite{Machacek:1983tz}. Thus for $N_f = 3$ and $N_H=1$ we can reproduce the numbers of equation \ref{bfunccoeffSM}.

An alternative approach to calculate these $b_i$s can be found in Ref.~\cite{Aitchison:2005cf}. For a $U(1)_Y$ gauge theory the coefficient $b_1$ is given by 
\be
\label{b1General}
b_1 = -\frac{2}{3}\sum_f Y_f^2 - \frac{1}{3}\sum_s Y_s^2
\ee
in which the fermionic matter particles have charges $Y_f$, the scalars have charges $Y_s$ and $Y$ is the normalized generator related to the hypercharge ($y$) as 
\be \label{Normalization}
Y^2 = \frac{3}{5} y^2,
\ee
and the factors 2/3 and 1/3 represent the contribution of fermion and scalar component respectively.
Corresponding $b$ coefficient formula for $SU(N)$ gauge theory is 
\be
\label{bNGeneral}
b_N = -\frac{11}{3}N - \frac{1}{3}n_f  - \frac{1}{6}n_s
\ee 
where $n_f$ is the number of left-handed fermions and $n_s$ is the number of complex scalars which couple to gauge bosons. Therefore in the SM, for $U(1)_Y$,
\bea
\label{b1SM}
(b_1)_{\mathrm{SM}} &=& -\frac{2}{3}\frac{3}{5}\sum_f y_f^2 - \frac{1}{3}\frac{3}{5}\sum_s y_s^2 \nonumber\\
&=& -\frac{2}{5}\cdot\tikzmark{s2}3\cdot\bigg[\tikzmark{a}3 y_{u_L}^2 + 3y_{d_L}^2 + 3y_{u_R}^2 + 3y_{d_R}^2 + y_{e_L}^2 +y_{\nu_L}^2 + y_{e_R}^2 \bigg] - \frac{1}{5}\big(\tikzmark{s1}2 y_{H}^2 \big ) \label{b1SM1}\\ 
\begin{tikzpicture}[remember picture,overlay]
\draw[<-] 
  ([shift={(2pt,-2pt)}]pic cs:a) |- ([shift={(14pt,-10pt)}]pic cs:a) 
  node[anchor=west] {$\scriptstyle \text{Color}$}; 
\end{tikzpicture} \nonumber\\ 
\begin{tikzpicture}[remember picture,overlay]
\draw[<-] 
  ([shift={(2pt,-2pt)}]pic cs:s1) |- ([shift={(14pt,-10pt)}]pic cs:s1) 
  node[anchor=west] {$\scriptstyle \text{Doublet}$}; 
\end{tikzpicture}
\begin{tikzpicture}[remember picture,overlay]
\draw[<-] 
  ([shift={(2pt,-2pt)}]pic cs:s2) |- ([shift={(16pt,-16pt)}]pic cs:s2) 
  node[anchor=west] {$\scriptstyle \text{Generation}$}; 
\end{tikzpicture}
&=& -\frac{2}{5}\cdot 3\cdot\bigg[3(\frac{1}{6})^2 + 3(\frac{1}{6})^2 + 3(-\frac{2}{3})^2 + 3(\frac{1}{3})^2 + (-\frac{1}{2})^2 + (-\frac{1}{2})^2 + 1 \big] - \frac{1}{5}\big[2(\frac{1}{2})^2\bigg]\nonumber \\
&=& -\frac{2}{5}\cdot 3\cdot\bigg[\frac{1}{6} + \frac{4}{3} + \frac{1}{3} + \frac{1}{2} + 1 \bigg] - \frac{1}{5}\cdot\frac{1}{2}\nonumber \\
&=& -\frac{2}{5}\cdot 3\cdot\frac{10}{3} - \frac{1}{10}\nonumber\\
&=& \boxed{- \frac{41}{10}}.
\eea
Similarly, for $SU(2)_L$,
\bea
\label{b2SM}
(b_2)_{\mathrm{SM}} &=& \frac{11}{3}\cdot 2 - \frac{1}{3}\cdot 12  - \frac{1}{6}\cdot 1 \nonumber \\
&=& \boxed{\frac{19}{6}}
\eea
where we have used the fact $n_f = 12$ (for each generation 1 doublet pair for quark and lepton, so for 3 generations there are total 6 pairs of left-handed fermions) and the Higgs is a doublet complex scalar only under $SU(2)_L$ gauge group. For $SU(3)_C$, only quarks carry the color and
\bea
\label{b3SM}
(b_3)_{\mathrm{SM}} &=& \frac{11}{3}\cdot 3 - \frac{1}{3}\cdot 12  - \frac{1}{6}\cdot 0 \nonumber \\
&=& \boxed{7}.
\eea
Thus we see that with a consistent difference in sign we can reproduce Eqn. \ref{bfunccoeffSM}.


\subsection{2-loop gauge $\b$-function coefficients}

For the SM, 2-loop $\b$-function coefficients are obtained as  following \cite{Machacek:1983tz}

\be
\label{BSMcoeffDerv}
\big(B_{ij}\big)_{\rm SM}= \begin{pmatrix}
0&0&0\cr 0 & \frac{136}{3}&0 \cr
0&0&102
\end{pmatrix} -n_f \begin{pmatrix}
\frac{19}{15}&\frac{3}{5}&\frac{44}{15}\cr \frac{1}{5} & \frac{49}{3}& 4 \cr \frac{11}{30}& \frac{3}{2}& \frac{76}{3}
\end{pmatrix} - \begin{pmatrix}
\frac{9}{50}&\frac{9}{10}&0 \cr \frac{3}{10} & \frac{13}{6}& 0 \cr 0 & 0 & 0
\end{pmatrix} = -\begin{pmatrix}
\frac{199}{50}&
\frac{27}{10}&\frac{44}{5}\cr \frac{9}{10} & \frac{35}{6}&12 \cr
\frac{11}{10}&\frac{9}{2}&-26
\end{pmatrix} 
\ee
for $n_f = 3$.
\noindent
Following equation \ref{genBcoeff}, we can derive the elements of each matrices above as follows. 
\bea
\label{SMB11}
(B_{11})_{\rm SM} &=& -\frac{34}{3}[C_2(G_1)]^2+\frac{1}{2\tikzmark{SMB111}}\bigg[4C_2(F_1)+\frac{20}{3}C_2(G_1)\bigg]S_2(F_1)\nonumber \\
&&\qquad +\tikzmark{SMB112}2\bigg[2C_2(S_1)+\frac{1}{3}C_2(G_1)\bigg]S_2(S_1)\nonumber \\
\begin{tikzpicture}[remember picture,overlay]
\draw[<-] 
  ([shift={(0pt,-0pt)}]pic cs:SMB111) |- ([shift={(10pt,-10pt)}]pic cs:SMB111) 
  node[anchor=west] {$\scriptstyle \text{Weyl~fermions}$};
  \draw[<-] 
  ([shift={(2pt,-2pt)}]pic cs:SMB112) |- ([shift={(12pt,-15pt)}]pic cs:SMB112) 
  node[anchor=west] {$\scriptstyle \text{Complex~scalar}$};  
\end{tikzpicture}
&=& 0+\frac{1}{2}\bigg[4\cdot\frac{3}{5}Y^2_{\rm fermions}+0\bigg]\frac{3}{5}Y^2_{\rm fermions}\cdot n_f+2\bigg[2\cdot\frac{3}{5}Y^2_{\rm scalars}+0\bigg]\frac{3}{5}Y^2_{\rm scalars}\label{SMB111}\\
&=& 0+\frac{18}{25}\cdot\bigg[Y^4_Q+Y^4_{\sbar u}+Y^4_{\sbar d}+Y^4_L+Y^4_{\sbar e}\bigg]+\frac{36}{25}Y^4_{\rm Higgs}\nonumber\\
&=& 0+\frac{18}{25}\cdot\bigg[(\tikzmark{SMB113}3\cdot \tikzmark{SMB114}2)\cdot(\frac{1}{6})^4+(3\cdot\tikzmark{SMB115}1)\cdot(-\frac{2}{3})^4+(3\cdot 1)\cdot(-\frac{1}{3})^4+2\cdot(-\frac{1}{2})^4 + 1 \bigg]\cdot n_f + \frac{36}{25}\big[\tikzmark{SMB116}2\cdot(\frac{1}{2})^4\bigg]\nonumber \\
 \begin{tikzpicture}[remember picture,overlay]
  \draw[<-] 
  ([shift={(2pt,-2pt)}]pic cs:SMB113) |- ([shift={(-12pt,-14pt)}]pic cs:SMB113) 
  node[anchor=east] {$\scriptstyle \text{Color}$}; 
  \draw[<-] 
  ([shift={(5pt,-2pt)}]pic cs:SMB114) |- ([shift={(15pt,-14pt)}]pic cs:SMB114) 
  node[anchor=west] {$\scriptstyle \text{Doublet}$}; 
  \draw[<-] 
  ([shift={(5pt,-2pt)}]pic cs:SMB115) |- ([shift={(15pt,-14pt)}]pic cs:SMB115) 
  node[anchor=west] {$\scriptstyle \text{Singlet}$}; 
  \draw[<-] 
  ([shift={(2pt,-2pt)}]pic cs:SMB116) |- ([shift={(-12pt,-14pt)}]pic cs:SMB116) 
  node[anchor=east] {$\scriptstyle \text{Doublet}$}; 
  \end{tikzpicture}
&=& 0+\bigg[\frac{1}{6^3}+\frac{16}{27}+\frac{1}{27}+\frac{1}{8}+1\bigg]\cdot n_f+\frac{36}{25}\cdot\frac{1}{8}\nonumber\\
&=&0+\bigg[\frac{1+128+8+27+216}{216}\bigg]\cdot n_f+\frac{9}{50}\nonumber\\
&=&\boxed{0+\frac{19}{15}\cdot n_f+\frac{9}{50}}
\eea
where in equation \ref{SMB111}, we have used \ref{CasGrp} and \ref{DynGrp} and thus we have recovered the 11 element of the matrix \ref{BSMcoeffDerv}. 
\noindent
Similarly,
\bea
\label{SMB12}
(B_{12})_{\rm SM} &=& -\frac{34}{3}\cdot 0+\frac{1}{2}\bigg[4C_2(F_2)+\frac{20}{3}\cdot 0\cdot 0\bigg]S_2(F_1)+2\bigg[2C_2(S_2)+\frac{1}{3}\cdot 0\cdot 0\bigg]S_2(S_1)\nonumber \\
&=& 0+\frac{1}{2}\cdot 4C_2(F_2)S_2(F_1)+2\cdot 2C_2(S_2)S_2(S_1)\nonumber \\
&=& 0+\frac{1}{2}\cdot 4\cdot\frac{3}{5}\big[Y_Q^2+Y_L^2\big]\cdot\frac{3}{4}+2\cdot 2\cdot\frac{3}{5}Y_{\rm Higgs}^2\cdot\frac{3}{4}\label{SMB121}
\\
&=& 0+\frac{9}{10}\big[3\cdot 2\cdot\big(\frac{1}{6}\big)^2+2.\big(\frac{1}{2}\big)^2\big]\cdot n_f+\frac{9}{5}\cdot 2\cdot \big(\frac{1}{2}\big)^2\nonumber\\
&=& \boxed{0+\frac{3}{5}\cdot n_f+\frac{9}{10}}
\eea
where in \ref{SMB121} we have used \ref{defC2} and \ref{DynGrp}.

\noindent
For 13 element calculation we need those multiplets which have color. Thus,
\bea
\label{SMB13}
(B_{13})_{\rm SM} &=& 0+\frac{1}{2}\cdot 4C_2(F_3)S_2(F_1)+2.2C_2(S_3)S_2(S_1)\nonumber \\
&=& 0+\frac{1}{2}\cdot 4\cdot\frac{3}{5}\big[Y_Q^2+Y_{\sbar u}^2+Y_{\sbar d}^2\big]\cdot\frac{4}{3}+3\cdot\frac{3}{5}Y_{\rm Higgs}^2\cdot 0
\\
&=& 0+\frac{8}{5}\big[3\cdot 2\cdot\big(\frac{1}{6}\big)^2+3\cdot\big(-\frac{2}{3}\big)^2+3\cdot\big(\frac{1}{3}\big)^2\big]\cdot n_f+0\nonumber\\
&=& 0+\frac{8}{5}\big[\frac{1}{6}+\frac{4}{3}+\frac{1}{3}\big]\cdot n_f+0\nonumber\\
&=& \boxed{0+\frac{44}{15}\cdot n_f+0}.
\eea

\bea
\label{SMB21}
(B_{21})_{\rm SM} &=& 0+\frac{1}{2}\cdot 4C_2(F_1)S_2(F_2)+2\cdot 2C_2(S_1)S_2(S_2)\nonumber \\
&=& 0+\frac{1}{2}\cdot 4\cdot\frac{1}{2}\cdot\frac{3}{5}\big[Y_Q^2+Y_{L}^2\big]+2\cdot 2\cdot\frac{1}{2}\cdot\frac{3}{5}Y_{\rm Higgs}^2 \nonumber\\
&=& 0+\frac{3}{5}\big[3\cdot\big(\frac{1}{6}\big)^2+1\cdot\big(-\frac{1}{2}\big)^2\big]\cdot n_f+\frac{6}{5}\big(\frac{1}{2}\big)^2\nonumber\\
&=& 0+\frac{3}{5}\big[\frac{1}{12}+\frac{1}{4}\big]\cdot n_f+\frac{3}{10}\nonumber\\
&=& \boxed{0+\frac{1}{5}\cdot n_f+\frac{3}{10}}.
\eea

\bea
\label{SMB22}
(B_{22})_{\rm SM} &=&-\frac{34}{3}[C_2(G_2)]^2+\frac{1}{2}\bigg[4C_2(F_2)|_{Q,L}+\frac{20}{3}C_2(G_2)\bigg]S_2(F_2)\nonumber \\&&\qquad +2\bigg[2C_2(S_2)|_{\rm Higgs}+\frac{1}{3}C_2(G_2)\bigg] S_2(S_2)\nonumber \\
&=&-\frac{34}{3}[2]^2+\frac{1}{2}\bigg[4\cdot\frac{3}{4}+\frac{20}{3}\cdot 2\bigg]\cdot\tikzmark{SMB221}4\cdot\frac{1}{2}\cdot n_f+2\bigg[2\cdot\frac{3}{4}+\frac{1}{3}\cdot 2\bigg].\frac{1}{2} \nonumber \\
\begin{tikzpicture}[remember picture,overlay]
  \draw[<-] 
  ([shift={(2pt,-2pt)}]pic cs:SMB221) |- ([shift={(12pt,-14pt)}]pic cs:SMB221) 
  node[anchor=west] {$\scriptstyle \text{(3+1) color multiplets}$}; 
\end{tikzpicture}  
 &=& \boxed{-\frac{136}{3}+\frac{49}{3}\cdot n_f+\frac{13}{6}}.
\eea
\noindent
For the calculation of $B_{23}$ element we only need to consider the contribution from the multiplet $Q$ as it is both doublet under $SU(2)$ and triplet under $SU(3)$ transformation.
\bea
\label{SMB23}
(B_{23})_{\rm SM} &=& 0+\frac{1}{2}\cdot 4C_2(F_3)S_2(F_2)+0\nonumber \\
&=& 0+\frac{1}{2}\cdot 4\cdot\bigg(\frac{4}{3}.\tikzmark{SMB231}3\bigg)\cdot\frac{1}{2}\cdot n_f+ 0 \nonumber\\
\begin{tikzpicture}[remember picture,overlay]
  \draw[<-] 
  ([shift={(2pt,-2pt)}]pic cs:SMB231) |- ([shift={(12pt,-14pt)}]pic cs:SMB231) 
  node[anchor=west] {$\scriptstyle \text{Color}$}; 
\end{tikzpicture} 
&=& \boxed{0+4\cdot n_f+0}.
\eea
\noindent
31 element calculation is similar as 13 element, with the interchange of the order of Casimir invariants and the Dynkin index. Therefore,
\bea
\label{SMB31}
(B_{31})_{\rm SM} &=& 0+\frac{1}{2}\cdot 4C_2(F_1)S_2(F_3)+2\cdot 2C_2(S_1)S_2(S_3)\nonumber \\
&=& 0+\frac{1}{2}\cdot 4\cdot\frac{3}{5}\big[Y_Q^2+Y_{\sbar u}^2+Y_{\sbar d}^2\big]\cdot\frac{1}{2}+0 \nonumber\\
&=& 0+\frac{3}{5}\cdot\tikzmark{SMB311}3\cdot\big[\tikzmark{SMB312}2\cdot\big(\frac{1}{6}\big)^2+\big(-\frac{2}{3}\big)^2+\big(\frac{1}{3}\big)^2\big]\cdot n_f+0\nonumber\\
\begin{tikzpicture}[remember picture,overlay]
  \draw[<-] 
  ([shift={(2pt,-2pt)}]pic cs:SMB311) |- ([shift={(-12pt,-14pt)}]pic cs:SMB311) 
  node[anchor=east] {$\scriptstyle \text{Color}$};
  \draw[<-] 
  ([shift={(2pt,-2pt)}]pic cs:SMB312) |- ([shift={(12pt,-14pt)}]pic cs:SMB312) 
  node[anchor=west] {$\scriptstyle \text{Doublet}$}; 
\end{tikzpicture}
&=& 0+\frac{9}{5}\big[\frac{1}{18}+\frac{5}{9}\big]\cdot n_f+0\nonumber\\
&=& \boxed{0+\frac{11}{10}\cdot n_f+0}.
\eea
Similar as the 23 element, the only contributing multiplet is $Q$ and 
\bea
\label{SMB32}
(B_{32})_{\rm SM} &=& 0+\frac{1}{2}\cdot 4 C_2(F_2)S_2(F_3)+0\nonumber \\
&=& 0+2\cdot\big(\frac{3}{4}\cdot 2\big)\frac{1}{2}\cdot n_f + 0\nonumber \\
&=& \boxed{0+\frac{3}{2}\cdot n_f+0}
\eea
Finally,
\bea
\label{SMB33}
(B_{33})_{\rm SM} &=&-\frac{34}{3}[C_2(G_3)]^2+\frac{1}{2}\bigg[4C_2(F_3)|_{Q,u^c,d^c}+\frac{20}{3}C_2(G_3)\bigg]S_2(F_3)\nonumber \\&&\qquad +2\bigg[2C_2(S_3)|_{\rm Higgs}+\frac{1}{3}C_2(G_3)\bigg] S_2(S_3)\nonumber \\
&=&-\frac{34}{3}\cdot[3]^2+\frac{1}{2}\bigg[4\cdot\frac{4}{3}+\frac{20}{3}\cdot 3\bigg](\tikzmark{SMB331}2+1+1)\cdot\frac{1}{2}\cdot n_f+0 \nonumber \\
\begin{tikzpicture}[remember picture,overlay]
  \draw[<-] 
  ([shift={(2pt,-2pt)}]pic cs:SMB331) |- ([shift={(12pt,-14pt)}]pic cs:SMB331) 
  node[anchor=west] {$\scriptstyle \text{Doublet}$};  
\end{tikzpicture}  
&=& -102+\bigg[\frac{16}{3}+20\bigg]\cdot n_f+0\nonumber\\
 &=& \boxed{-102+\frac{76}{3}\cdot n_f+0}.
\eea


\section{$\b$-functions of MSSM Gauge coupling RGEs (SUSY Scenario)}
\label{MSSMRGEs}

The MSSM particle content is summarized in Table \ref{tab:chiral}.
 
\begin{table}[H]
\begin{center}
\begin{tabular}{|c|c|c|c|c|}
\hline
\multicolumn{2}{|c|}{Names} 
& spin 0 & spin 1/2 & $SU(3)_C ,\, SU(2)_L ,\, U(1)_Y$
\\  \hline\hline
squarks, quarks & $Q$ & $({\stilde u}_L\>\>\>{\stilde d}_L )$&
 $(u_L\>\>\>d_L)$ & $(\>{\bf 3},\>{\bf 2}\>,\>{1\over 6})$
\\
($\times 3$ families) & $\sbar u$
&${\stilde u}^*_R$ & $u^\dagger_R$ & 
$(\>{\bf \overline 3},\> {\bf 1},\> -{2\over 3})$
\\ & $\sbar d$ &${\stilde d}^*_R$ & $d^\dagger_R$ & 
$(\>{\bf \overline 3},\> {\bf 1},\> {1\over 3})$
\\  \hline
sleptons, leptons & $L$ &$({\stilde \nu}\>\>{\stilde e}_L )$&
 $(\nu\>\>\>e_L)$ & $(\>{\bf 1},\>{\bf 2}\>,\>-{1\over 2})$
\\
($\times 3$ families) & $\sbar e$
&${\stilde e}^*_R$ & $e^\dagger_R$ & $(\>{\bf 1},\> {\bf 1},\>1)$
\\  \hline
Higgs, higgsinos &$H_u$ &$(H_u^+\>\>\>H_u^0 )$&
$(\stilde H_u^+ \>\>\> \stilde H_u^0)$& 
$(\>{\bf 1},\>{\bf 2}\>,\>+{1\over 2})$
\\ &$H_d$ & $(H_d^0 \>\>\> H_d^-)$ & $(\stilde H_d^0 \>\>\> \stilde H_d^-)$& 
$(\>{\bf 1},\>{\bf 2}\>,\>-{1\over 2})$
\\  \hline
\end{tabular}
\caption{Chiral supermultiplets in the Minimal Supersymmetric Standard Model.
The spin-$0$ fields are complex scalars, and the spin-$1/2$ fields are 
left-handed two-component Weyl fermions.\label{tab:chiral}}
\vspace{-0.6cm}
\end{center}
\end{table}

The two-loop renormalization group equations for the gauge couplings in MSSM are
\begin{eqnarray}
(4\pi)^2\frac{d}{dt}~ g_i &=& g_i^3 b_i 
 +\frac{g_i^3}{(4\pi)^2}
\left[~ \sum_{j=1}^3 B_{ij}  g_j^2-\sum_{\alpha=u,d,e}d_i^\alpha
{\rm Tr}\left( y^{\alpha \dagger}y^{\alpha}\right) \right] ~,~\,
\label{SUSYgauge}
\end{eqnarray}
where the beta-function coefficients for 1-loop gauge, 2-loop gauge and 2-loop Yukawa couplings are 
\be
\label{MSSMbcoeff}
b_{\mathrm{MSSM}}=\left(\frac{33}{5},1,-3\right), 
\ee
\be
\label{MSSMBcoeff}
\big(B_{ij}\big)_{\rm MSSM}=\begin{pmatrix}\frac{199}{25}&~~
\frac{27}{5}& ~~\frac{88}{5}\cr \frac{9}{5} & ~~25& ~~24 \cr
\frac{11}{5}& ~~9& ~~14 \end{pmatrix},
\ee
and 
\begin{eqnarray}
\label{MSSMYukcoeff}
&&\big(d^u\big)_{\rm{MSSM}}=\left(\frac{26}{5},6,4\right) ~,~
\big(d^d\big)_{\rm{MSSM}}=\left(\frac{14}{5},6,4\right) ~,~
\big(d^e\big)_{\rm{MSSM}}=\left(\frac{18}{5},2,0\right)  
\end{eqnarray}
respectively.


\subsection{1-loop gauge $\b$-function coefficients}

To derive \ref{MSSMbcoeff}, the equation \ref{bNGeneral} needs to be changed because now there are some new particles whose contributions are significant. Following are those contributions:
\begin{itemize}
\item Contributions of gauginos to $SU(N)$ gauge group is $-\frac{2N}{3}$ and to $U(1)$ is zero.
\item Contributions of squarks and sleptons or combinedly sfermions (scalars) to $SU(N)$ gauge group is $-\frac{1}{6}n_s$  and to $U(1)$ is $-\frac{3}{5}\sum_s y_s^2$.
\item Contributions of Higgsinos (fermions) to $SU(N)$ gauge group is $-\frac{1}{3}n_f$  and to $U(1)$ is $-\frac{3}{5}\sum_f y_f^2$.
\end{itemize}
Therefore following equation \ref{b1General}, the $U(1)_Y$ coefficient for the MSSM case is:
\bea
\label{b1MSSM}
(b_1)_{\mathrm{MSSM}} &=& (b_1)_{\mathrm{SM}} -\frac{1}{3}\cdot\frac{3}{5}\sum_{s=\mathrm{Scalars}} y_s^2 - \frac{2}{3}\cdot\frac{3}{5}\sum_{f=\mathrm{Fermions}} y_f^2 \\
&=& -\frac{41}{10} - \frac{1}{5}\cdot\big(\tikzmark{aa}3\cdot\frac{10}{3} +\tikzmark{a1} 2\cdot\frac{1}{4}\big) - \frac{2}{5}\cdot\tikzmark{a2}4\cdot\frac{1}{4}\label{b1MSSM1}\\
\begin{tikzpicture}[remember picture,overlay]
\draw[<-] 
  ([shift={(2pt,-2pt)}]pic cs:aa) |- ([shift={(-10pt,-11pt)}]pic cs:aa) 
  node[anchor=east] {$\scriptstyle \text{Color}$};
  \draw[<-] 
  ([shift={(2pt,-2pt)}]pic cs:a1) |- ([shift={(12pt,-15pt)}]pic cs:a1) 
  node[anchor=west] {$\scriptstyle \text{Doublet}$};  
  \draw[<-] 
  ([shift={(2pt,-2pt)}]pic cs:a2) |- ([shift={(12pt,-15pt)}]pic cs:a2) 
  node[anchor=west] {$\scriptstyle \text{2~Doublets}$};  
\end{tikzpicture} 
&=& -\frac{41}{10} - \frac{1}{5}\cdot\frac{21}{2} - \frac{2}{5}\nonumber\\
&=& \boxed{-\frac{33}{5}}.
\eea
Points about Eqn. \ref{b1MSSM1}
\begin{itemize}
\item First term is the SM contribution coming only from the fermions of 3 colors and generations and from the single Higgs.
\item Second term is a combined contribution of both, the additional scalar partners of quarks and leptons introduced in the multiplet and the additional scalar Higgs introduced in the MSSM.
\item Since there are now two Higgs doublet in the MSSM, there are two additional fermionic partners to them, Higgsinos. Their contribution is given by this last term. 
\end{itemize}
Following equation \ref{bNGeneral}, the $SU(2)_L$ coefficient for the MSSM is
\bea
\label{b2MSSM}
(b_2)_{\mathrm{MSSM}} &=& (b_2)_{\mathrm{SM}} -\frac{2N}{3} -\frac{1}{3}n_f -\frac{1}{6}n_s\\
&=& \frac{19}{6}-\frac{4}{3}-\frac{1}{3}\cdot 2-\frac{1}{6}\cdot(12+1)\label{b2MSSM2}\\
&=& \boxed{-1}.
\eea
Points about Eqn. \ref{b2MSSM2}
\begin{itemize}
\item First term is the usual SM contribution.
\item For $SU(2)$ case, $N=2$ in the second term.
\item Two Higgsinos are the extra left-handed fermions present in the MSSM, so the presence of 2 in the third term.
\item The additional scalar partners of quarks and leptons introduced in the multiplet (1 doublet each for squarks and sleptons and 3 generations, so total 6 doublets and thus total 12 sfermions under $SU(2)_L$) and one additional scalar Higgs introduced in the MSSM.
\end{itemize}

\noindent
Similarly, following equation \ref{bNGeneral}, the MSSM $SU(3)_C$ coefficient is
\bea
\label{b3MSSM}
(b_3)_{\mathrm{MSSM}} &=& (b_3)_{\mathrm{SM}} -\frac{2N}{3} -\frac{1}{3}n_f -\frac{1}{6}n_s\\
&=& 7 - 2 -(\frac{1}{3}\cdot 0)-(\frac{1}{6}\cdot 12)\label{b3MSSM2}\\
&=& \boxed{3}.
\eea
Points about Eqn. \ref{b3MSSM2}
\begin{itemize}
\item First term is the usual SM contribution.
\item For $SU(3)$ case, $N=3$ in the second term.
\item Higgsinos don't contribute as they are singlet under the $SU(3)$ representation.
\item For each generation we need to consider only the squarks contributions. There are four (1 doublet and 2 singlets) $SU(3)$ fundamental representations. Thus for 3 generations there are total 12 additional sfermions (scalars). Again being singlet under $SU(3)$, there is no contribution from the extra Higgs doublet, introduced in the MSSM. 
\end{itemize}

\noindent
Thus we reproduced the equation \ref{MSSMbcoeff}, again with a consistent difference in sign.


\subsection{2-loop gauge $\b$-function coefficients}

2-loop $\b$-functions for gauge couplings is given by \cite{Martin:1993zk}

\be \label{2loopbetaMSSM}
\b_g^(2) = g^5 \big\{-6[C(G)]^2 + 2C(G)S(R) + 4S(R)C(R)\big\} - g^3 Y^{ ijk}Y_{ijk}C(k)/d(G) 
\ee
where the second term is the Yukawa coupling contribution and  $Y_{ijk} = (Y^{ijk})^*$, $C(G)$ is the Casimir invariant of the group, $S(R)$ is the Dynkin index summed over all chiral multiplets and $S(R)C(R)$ is the sum of the Dynkin indices weighted by the quadratic Casimir group theory invariant $C(R)$ for the representation $R$. $C(R)=C_R(i)$ for the superfield $\Phi_i$, defined in terms of the Lie algebra generators $T^a$
by 
\be
(T^aT^a)_i{}^{j}= C_R(i) \delta_i^j .
\label{eq:defCasimir}
\ee
The Dynkin index $\rm S(R)$ for an irreducible representation $R$ is defined by equation \ref{eq:defDynkin} and also the relation between the Casimir invariant and the Dynkin index is given by \ref{CasDyn}. Explicitly, the Casimir invariants for the MSSM supermultiplets in different representation (in the superscripts) are given by equations \ref{defC1}-\ref{defC3} with $H$ replaced by $H_u$ and $H_d$.

Now using Equations \ref{defC1} - \ref{DynGrp} let us calculate the elements of $B_{\mathrm{MSSM}}$ from the first term of Equation \ref{2loopbetaMSSM}. Let us just define the elements of $B_{\mathrm{MSSM}}$ as 

\be \label{MSSMBMat}
B_{\mathrm{MSSM}}=\begin{pmatrix}\frac{199}{25}&
~~\frac{27}{5}& ~~\frac{88}{5}\cr \frac{9}{5} & ~~25& ~~24 \cr
\frac{11}{5}& ~~9& ~~14 \end{pmatrix}=\begin{pmatrix}B_{YY}&~~
B_{YL}& ~~B_{YC}\cr B_{LY} & ~~B_{LL}& ~~B_{LC} \cr
B_{CY} & ~~B_{CL} & ~~B_{CC} \end{pmatrix}
\ee
where $Y, L, C$ represent $U(1)_Y, SU(2)_L$ and $SU(3)_C$ respectively. The ordering is important as will be clear in the following calculations.

\bea \label{B11MSSM}
\left(B_{YY}\right)_{\rm MSSM} &=& -6\cdot 0+2\cdot 0+4 \sum S(R)C(R)~~[\mathrm{for}~ U(1), C_2(G)=0] \nonumber\\
 &=& 4 \sum_{\Phi_i} \big (\frac{3}{5}Y_{\Phi_i}^2 \big)\big (\frac{3}{5}Y_{\Phi_i}^2 \big)~~[\Phi_i=\mathrm{All~MSSM~ supermultiplets}]\nonumber\\ 
 &=& \frac{36}{25} \sum_{\Phi_i}Y_{\Phi_i}^4 \nonumber\\ 
 &=& \frac{36}{25}\bigg[\tikzmark{YY1}3\bigg\{(\tikzmark{YY2}2\cdot\tikzmark{YY3}3)\cdot(\frac{1}{6})^4+3\cdot(-\frac{2}{3})^4+3\cdot(-\frac{1}{3})^4+2\cdot(-\frac{1}{2})^4\bigg\}+(2\cdot 2)\cdot(\frac{1}{2})^4\bigg]\nonumber\\ 
 \begin{tikzpicture}[remember picture,overlay]
\draw[<-] 
  ([shift={(2pt,-2pt)}]pic cs:YY1) |- ([shift={(-10pt,-12pt)}]pic cs:YY1) 
  node[anchor=east] {$\scriptstyle \text{Generation}$}; 
  \draw[<-] 
  ([shift={(2pt,-2pt)}]pic cs:YY2) |- ([shift={(24pt,-12pt)}]pic cs:YY2) 
  node[anchor=west] {$\scriptstyle \text{Doublet}$}; 
  \draw[<-] 
  ([shift={(5pt,-2pt)}]pic cs:YY3) |- ([shift={(20pt,-18pt)}]pic cs:YY3) 
  node[anchor=west] {$\scriptstyle \text{Color}$}; 
  \end{tikzpicture}
 &=& \frac{36}{25}.\frac{398}{72}\nonumber\\ 
 &=& \boxed{\frac{199}{25}}
\eea

\bea \label{B12MSSM}
\left(B_{YL}\right)_{\rm MSSM} &=& 4\sum C_2(F_2)S_2(F_1)\nonumber\\ 
 &=& 4 \cdot\frac{3}{4}\cdot\sum_{\Phi_i} \left(\frac{3}{5}Y_{\Phi_i}^2\right)~~[\Phi_i=Q,L,H_u,H_d~(\mathrm{doublets~under~}SU(2))]\nonumber\\ 
 &=& \frac{9}{5} \big[Y_{Q}^2+Y_{L}^2+Y_{H_u}^2+Y_{H_d}^2\big]\nonumber \\ 
 &=& \frac{9}{5}\bigg[\tikzmark{YL1}3\cdot\bigg\{(\tikzmark{YL2}2\cdot\tikzmark{YL3}3)\cdot(\frac{1}{6})^2+2\cdot(-\frac{1}{2})^2\bigg\}+2\cdot(-\frac{1}{2})^2+2\cdot(\frac{1}{2})^2\bigg]\nonumber\\ 
 \begin{tikzpicture}[remember picture,overlay]
\draw[<-] 
  ([shift={(2pt,-2pt)}]pic cs:YL1) |- ([shift={(-10pt,-12pt)}]pic cs:YL1) 
  node[anchor=east] {$\scriptstyle \text{Generation}$}; 
  \draw[<-] 
  ([shift={(2pt,-2pt)}]pic cs:YL2) |- ([shift={(24pt,-16pt)}]pic cs:YL2) 
  node[anchor=west] {$\scriptstyle \text{Doublet}$};  
\draw[<-] 
  ([shift={(2pt,-2pt)}]pic cs:YL3) |- ([shift={(14pt,-10pt)}]pic cs:YL3) 
  node[anchor=west] {$\scriptstyle \text{Color}$};  
\end{tikzpicture}
 &=& \boxed{\frac{27}{5}}
\eea

\bea \label{B13MSSM}
\left(B_{YC}\right)_{\rm MSSM} &=& 4\sum C_2(F_3)S_2(F_1)\nonumber\\ 
 &=& 4\cdot \frac{4}{3}\cdot\sum_{\Phi_i}\big(\frac{3}{5}Y_{\Phi_i}^2\big)~~[\Phi_i=Q,\bar{u},\bar{d}~(\mathrm{triplets~under~}SU(3))]\nonumber\\ 
 &=& \frac{16}{5} \big[Y_{Q}^2+Y_{\bar{u}}^2+Y_{\bar{d}}^2\big]\nonumber \\ 
 &=& \frac{16}{5}\bigg[(\tikzmark{YC1}3\cdot\tikzmark{YC2}3)\cdot\bigg\{2\cdot(\frac{1}{6})^2+(-\frac{2}{3})^2+(\frac{1}{3})^2\bigg\}\bigg]\nonumber\\ 
 \begin{tikzpicture}[remember picture,overlay]
\draw[<-] 
  ([shift={(2pt,-2pt)}]pic cs:YC1) |- ([shift={(-10pt,-12pt)}]pic cs:YC1) 
  node[anchor=east] {$\scriptstyle \text{Generation}$}; 
  \draw[<-] 
  ([shift={(2pt,-2pt)}]pic cs:YC2) |- ([shift={(24pt,-14pt)}]pic cs:YC2) 
  node[anchor=west] {$\scriptstyle \text{Color}$}; 
  \end{tikzpicture}
 &=& \boxed{\frac{88}{5}}
\eea

\bea \label{B21MSSM}
\left(B_{LY}\right)_{\rm MSSM} &=& 4\sum C_2(F_1)S_2(F_2)\nonumber\\ 
 &=& 4\cdot \frac{1}{2}\cdot\sum_{\Phi_i} \frac{3}{5}Y_{\Phi_i}^2 ~~[\Phi_i=Q,L,H_u,H_d~(\mathrm{doublets~under~}SU(2))]\nonumber\\ 
 &=& \frac{6}{5} \big[Y_{Q}^2+Y_{L}^2+Y_{H_u}^2+Y_{H_d}^2\big]\nonumber \\ 
 &=& \frac{16}{5}\bigg[(3\cdot 3)\cdot(\frac{1}{6})^2+(1\cdot 3)\cdot(-\frac{2}{3})^2+1\cdot(\frac{1}{2})^2+1\cdot(-\frac{1}{2})^2\bigg]\nonumber \\ 
 &=& \boxed{\frac{9}{5}}
\eea

\bea \label{B22MSSM}
\left(B_{LL}\right)_{\rm MSSM} &=& -6(2)^2+(2\cdot 2)\sum_{\Phi_i}(\frac{1}{2})\cdot\Phi_i+4\cdot \frac{1}{2}\sum_{\Phi_i}(\frac{3}{4})\cdot\Phi_i \nonumber\\ 
 &=& -24 + (2\cdot 2)\cdot(\frac{1}{2}\times14)+4\cdot\frac{1}{2}(\frac{3}{4}\times14)\label{B22MSSM1}\\ 
 &=& \boxed{25}
\eea
where in Equation \ref{B22MSSM1},
\bea
\label{multiplets}
14 &=& \left( n_Q + n_L + n_{H_u} + n_{H_d}\right)\nonumber\\ 
14 &=& (\tikzmark{Mul1}3\times \tikzmark{Mul2}3+1\times \tikzmark{Mul3}3+1\times 1+1\times 1)
\begin{tikzpicture}[remember picture,overlay]
\draw[<-] 
  ([shift={(2pt,-2pt)}]pic cs:Mul1) |- ([shift={(-10pt,-12pt)}]pic cs:Mul1) 
  node[anchor=east] {$\scriptstyle \text{Generation}$}; 
  \draw[<-] 
  ([shift={(2pt,-2pt)}]pic cs:Mul2) |- ([shift={(12pt,-12pt)}]pic cs:Mul2) 
  node[anchor=west] {$\scriptstyle \text{Color}$}; 
  \draw[<-] 
  ([shift={(2pt,-2pt)}]pic cs:Mul3) |- ([shift={(20pt,-12pt)}]pic cs:Mul3) 
  node[anchor=west] {$\scriptstyle \text{Generation}$}; 
  \end{tikzpicture}
\eea 
is the total number of multiplets from $Q, L, H_u$ and $H_d$. Similarly,

\bea 
\left(B_{CC}\right)_{\rm MSSM} &=& -6(3)^2+(2\cdot 3)\sum_{\Phi_i}(\frac{1}{2})\cdot\Phi_i+4\cdot\frac{1}{2}\sum_{\Phi_i}(\frac{4}{3})\cdot\Phi_i \nonumber\\ 
 &=& -54 + 2\cdot 3\cdot(\frac{1}{3}\times 12)+4\cdot\frac{1}{2}\cdot(\frac{4}{3}\times12)\label{B33MSSM1}\\ 
 &=& \boxed{14}
 \label{B33MSSM}
\eea
where in Equation \ref{B33MSSM1}, 
\be
12 = \tikzmark{Mul4}3\cdot\left(2+1+1\right)
\begin{tikzpicture}[remember picture,overlay]
\draw[<-] 
  ([shift={(2pt,-2pt)}]pic cs:Mul4) |- ([shift={(10pt,-12pt)}]pic cs:Mul4) 
  node[anchor=west] {$\scriptstyle \text{Generation}$};
  \end{tikzpicture}
\ee
is again the total number of multiplets from $Q, \bar{u}, \bar{d}$.

\bea 
\label{B23MSSM1}
\left(B_{LC}\right)_{\rm MSSM} &=& 4\sum C_2(F_3)S_2(F_2) \nonumber\\
 &=& 4\cdot \frac{1}{2}\cdot\frac{4}{3}\cdot\sum_{\Phi_i} \Phi_i ~~[\Phi_i=Q,\bar{u},\bar{d}]\nonumber\\ 
 &=& 4\cdot\frac{1}{2}\cdot\frac{4}{3}\cdot(3\times3)\nonumber\\
 &=& \boxed{24}
\eea

\bea 
\left(B_{CY}\right)_{\rm MSSM} &=& 4\sum C_2(F_1)S_2(F_3)\nonumber\\ 
 &=& 4\cdot\frac{1}{2}\cdot\sum_{\Phi_i}\frac{3}{5}Y_{\Phi_i}^2~~[\Phi_i=Q,\bar{u},\bar{d}~(\mathrm{triplets~under~}SU(3))]\nonumber\\ 
 &=& \frac{6}{5}\cdot\big[(2\cdot 3)\cdot\big(\frac{1}{6}\big)^2+(1\cdot 3)\cdot\big(-\frac{2}{3}\big)^2+(1\cdot 3)\cdot\big(\frac{1}{3}\big)^2\big]\\ \label{B31MSSM1}
 &=& \boxed{\frac{11}{5}}
\eea

\bea 
\label{B32MSSM1}
\left(B_{CL}\right)_{\rm MSSM} &=& 4\sum C_2(F_2)S_2(F_3)\nonumber\\ 
 &=& 4\cdot\frac{1}{2}\cdot\frac{3}{4}\cdot\sum_{\Phi_i} \Phi_i ~~[\Phi_i=Q,L,H_u,H_d]\nonumber\\ 
 &=& 4\cdot\frac{1}{2}\cdot\frac{3}{4}\cdot(2\times3)\\ 
 &=& \boxed{9}
\eea


\section{$\beta$-function coefficients for $\uonep$ particles (SUSY Scenario)}
\label{UpRGEs}

Adding an extra $\uonep$ gauge group along with the usual SM gauge groups is one of the simplest and best motivated way to extend the MSSM. For the purpose of this note, let us take the example of this Ref. \cite{deBlas:2009vx}. To solve the $\mu$ problem they have introduced a SM singlet superfield $S$ which is charged under $\uonep$ and to cancel the anomalies in the model, 3 pairs of colored, $SU(2)_L$ singlet exotics $D, D^c$ with hypercharge $Y_D = -\frac{1}{3}$ and $Y_{D^c} = \frac{1}{3}$ and 2 pairs of uncolored $SU(2)_L$ singlet exotics $E, E^c$ with hypercharge $Y_E = −1$ and $Y_{E^c} = 1$ were introduced. Collecting the new particle's charges from the text, the quantum numbers of these new particles in $SU(3)_C\times SU(2)_L\times U(1)_Y\times \uonep $ representation become

\begin{table}[H]
\begin{center}
\begin{tabular}{|c|c|}
\hline
Names  & $SU(3)_C ,\, SU(2)_L ,\, U(1)_Y,\, \uonep$
\\  \hline\hline
 $D$ & $(\>{ 3},\>{ 1}\>,\>-{1\over 3},\>\frac{4}{5})$
\\ \hline
 $D^c$ & $(\>{ 3},\>{ 1}\>,\>{1\over 3},\>-\frac{1}{5})$
\\ \hline
 $E$ & $(\>{ 1},\>{ 1}\>,\> -1,\>\frac{9}{5})$
\\ \hline
 $E^c$ & $(\>{ 1},\>{ 1}\>,\> 1,\>-\frac{6}{5})$
\\ \hline
$S$ & $(\>{ 1},\>{ 1}\>,\> 0,\>-\frac{3}{5})$
\\ \hline
\end{tabular}
\caption{New particles under the $\uonep$ model.\label{tab:uonep}}
\vspace{-0.6cm}
\end{center}
\end{table}

The superpotential is given by
\begin{eqnarray}
\label{U1ppot}
W&=& y_u  {H}_u {Q} {u}^c + y_d  {H}_d {Q} {d}^c + y_e  {H}_d {L} {e}^c+y_\nu  {H}_u {L} {\nu}^c
\\\nonumber
&+&\lambda  {S}  {H}_u  {H}_d+ y_D\, S\left( \sum_{i=1}^{3}  D_i D_i^c \right) +y_E\, S\left (\sum_{j=1}^{2} E_j E_j^c\right).
\end{eqnarray}

\noindent
The presence of these new particles will have significant effects on the corresponding RGEs and thus it will be interesting to see how we can derive the expressions summarized in the Appendix of Ref. \cite{deBlas:2009vx}. Let us calculate those.

\subsection{1-loop gauge $\b$ function coefficients}

The contribution of these particles to the 1-loop $\b$ function is 
\bea
b_{\mathrm{\uonep}} &=& \frac{3}{5}\big[Y_S^2+Y_{D,D^c}^2+Y_{E,E^c}^2\big]\qquad\qquad\qquad\qquad\nonumber \\ 
&=& \frac{3}{5}\big[0+(\tikzmark{up1L1}2\cdot\tikzmark{up1L2}3\cdot\tikzmark{up1L3}3\cdot)\cdot(\frac{1}{3})^2+(\tikzmark{up1L4}2\cdot 1\cdot\tikzmark{up1L5}2)\cdot 1\big]\\
\begin{tikzpicture}[remember picture,overlay]
\draw[<-] 
  ([shift={(2pt,-2pt)}]pic cs:up1L1) |- ([shift={(-10pt,-13pt)}]pic cs:up1L1) 
  node[anchor=east] {$\scriptstyle \text{D,$\rm D^c$}$};
  \draw[<-] 
  ([shift={(4pt,-2pt)}]pic cs:up1L2) |- ([shift={(14pt,-15pt)}]pic cs:up1L2) 
  node[anchor=west] {$\scriptstyle \text{Color}$};
  \draw[<-] 
  ([shift={(5pt,-2pt)}]pic cs:up1L3) |- ([shift={(30pt,-20pt)}]pic cs:up1L3) 
  node[anchor=west] {$\scriptstyle \text{Pairs}$};
  \draw[<-] 
  ([shift={(2pt,-2pt)}]pic cs:up1L4) |- ([shift={(10pt,-12pt)}]pic cs:up1L4) 
  node[anchor=west] {$\scriptstyle \text{E,$\rm E^c$}$};
  \draw[<-] 
  ([shift={(2pt,-2pt)}]pic cs:up1L5) |- ([shift={(25pt,-18pt)}]pic cs:up1L5) 
  node[anchor=west] {$\scriptstyle \text{Pairs}$};  
\end{tikzpicture} 
&=& \boxed{\frac{18}{5}}.
\eea
They all are singlet under $SU(2)$ transformation, so no contribution from them towards $g_2$ and for the such as the one considered in that reference, $g_3$ vanishes at 1-loop. Finally the contribution of the $\uonep$ to the 1-loop $\b$ function is just related to there charges under $\uonep$, $\mathrm{Tr}[Q_i^2]$. Therefore adding with the MSSM contribution \ref{MSSMbcoeff}

\bea
\label{uonepbcoeff}
b_{\mathrm{New}}=b_{\mathrm{MSSM}}+b_{\mathrm{\uonep}}&=&\left(\frac{33}{5},1,0\right) +\left(\frac{18}{5},0,0,\mathrm{Tr}[Q_i^2]\right) \nonumber\\
&=&\left(\frac{51}{5},1,0,\mathrm{Tr}[Q_i^2]\right).
\eea

\subsection{2-loop gauge $\b$ function coefficients}

Clearly due to the presence of an extra gauge group \ref{MSSMBMat} would be modified as
\be \label{UonepBMat}
B_{\mathrm{New}}=\begin{pmatrix}B_{YY}&
~~B_{YL}& ~~B_{YC}& ~~B_{YP}\cr B_{LY} & ~~B_{LL}& ~~B_{LC} & ~~B_{LP} \cr
B_{CY} & ~~B_{CL} & ~~B_{CC} & ~~B_{CP} \cr
B_{PY} & ~~B_{PL} & ~~B_{PC} & ~~B_{PP}  \end{pmatrix}_{\mathrm{New}}=\setlength\arraycolsep{2pt}
\begin{pmatrix}\frac{351}{25}&\frac{27}{5}
& 24 & \frac{12}{5}\mathrm{Tr}[Y_i^2Q_i^2]\cr \frac{9}{5} & 25 & 24 & (B_{LP})_{\mathrm{New}} \cr
3 & 9 & 48 & (B_{CP})_{\mathrm{New}} \cr
\frac{12}{5}\mathrm{Tr}[Y_i^2Q_i^2] & (B_{PL})_{\mathrm{New}}  & (B_{PC})_{\mathrm{New}} & \mathrm{Tr}[Q_i^4]  \end{pmatrix}
\ee
where
\bea
(B_{LP})_{\mathrm{New}} &=& 2\big[Q_{H_d}^2+Q_{H_u}^2+3(Q_L^2+3Q_Q^2)\big]\\
(B_{CP})_{\mathrm{New}} &=& 6\big[2Q_Q^2+Q_{u^c}^2+Q_D^2+Q_{D^c}^2\big]\\
(B_{PL})_{\mathrm{New}} &=& 6\big[2Q_{H_u}^2+Q_{H_d}^2+3(Q_L^2+Q_Q^2)\big]\\
(B_{PC})_{\mathrm{New}} &=& 48\big[2Q_Q^2+Q_{u^c}^2+Q_{d^c}^2+Q_D^2+_{D^c}^2\big]
\eea

Now, let us calculate the terms of \ref{UonepBMat} explicitly and derive the RGEs. Following Eqn. \ref{2loopbetaMSSM}, the contribution to the 2-loop $\b$-function coefficients due to the $\uonep$ gauge group are the following:

\bea \label{BYYUp}
(B_{YY})_{\mathrm{New}} &=& 4\sum S(R)C(R)\nonumber\\ 
 &=& 4\cdot (\frac{3}{5})^2\sum_{\Phi_i}Y_{\Phi_i}^4 ~~[\Phi_i=\mathrm{All~multiplets}]\nonumber\\ 
 &=& \frac{36}{25} \big[\left(\tikzmark{b}3\cdot\tikzmark{c}3\cdot\tikzmark{d}2\cdot\right)(\frac{1}{6})^4+(3\cdot 3)\cdot(-\frac{2}{3})^4+(3\cdot 3)\cdot(\frac{1}{3})^4+(3\cdot 2)\cdot(-\frac{1}{2})^4+3+2.(\frac{1}{2})^4 \label{BYYUp1}\\
\begin{tikzpicture}[remember picture,overlay]
\draw[<-] 
  ([shift={(2pt,-2pt)}]pic cs:b) |- ([shift={(-10pt,-12pt)}]pic cs:b) 
  node[anchor=east] {$\scriptstyle \text{Color}$}; 
\draw[<-] 
  ([shift={(2pt,-2pt)}]pic cs:c) |- ([shift={(14pt,-10pt)}]pic cs:c) 
  node[anchor=west] {$\scriptstyle \text{Generation}$}; 
  \draw[<-] 
  ([shift={(5pt,-2pt)}]pic cs:d) |- ([shift={(24pt,-18pt)}]pic cs:d) 
  node[anchor=west] {$\scriptstyle \text{Doublet}$}; 
\end{tikzpicture}
&&+2\cdot(\frac{-1}{2})^4+(3\cdot 3\cdot 2)\cdot(\frac{-1}{3})^4+(2\cdot 2)\cdot 1^4\big]\nonumber \\ 
 &=& \frac{36}{25}\bigg[7+\frac{1}{4}+\frac{3}{9}+\frac{16}{9}+\frac{3}{8}+\frac{1}{72}\bigg]\nonumber\\ 
 &=& \boxed{\frac{351}{25}}
\eea
$B_{YL}$ is calculated the same way as for the MSSM, in equation \ref{B12MSSM}.
\bea \label{BYCUp}
(B_{YC})_{\mathrm{New}} &=& 4\bigg(\frac{3}{5}\sum_{\Phi_i}Y_{\Phi_i}\bigg)\frac{4}{3}\nonumber\\ 
 &=& \frac{16}{5}\cdot\tikzmark{upYC1}3\bigg[(\tikzmark{upYC2}3)\cdot\tikzmark{upYC3}2(\frac{1}{6})^2+3\cdot(-\frac{2}{3})^2+3\cdot(\frac{1}{3})^2+(\tikzmark{upYC4}3\cdot\tikzmark{upYC5}2)\cdot(\frac{1}{3})^2\bigg]\nonumber\\ 
 &=& \frac{48}{5} \bigg[\frac{1}{6}+\frac{4}{3}+\frac{1}{3}+\frac{2}{3}\bigg]\nonumber\\
\begin{tikzpicture}[remember picture,overlay]
\draw[<-] 
  ([shift={(2pt,-2pt)}]pic cs:upYC1) |- ([shift={(-10pt,-12pt)}]pic cs:upYC1) 
  node[anchor=east] {$\scriptstyle \text{Color}$}; 
  \draw[<-] 
  ([shift={(2pt,-2pt)}]pic cs:upYC2) |- ([shift={(20pt,-18pt)}]pic cs:upYC2) 
  node[anchor=west] {$\scriptstyle \text{Generation}$}; 
\draw[<-] 
  ([shift={(2pt,-2pt)}]pic cs:upYC3) |- ([shift={(14pt,-10pt)}]pic cs:upYC3) 
  node[anchor=west] {$\scriptstyle \text{Doublet}$}; 
  \draw[<-] 
  ([shift={(2pt,-2pt)}]pic cs:upYC4) |- ([shift={(-10pt,-12pt)}]pic cs:upYC4) 
  node[anchor=east] {$\scriptstyle \text{Pairs}$}; 
  \draw[<-] 
  ([shift={(2pt,-2pt)}]pic cs:upYC5) |- ([shift={(20pt,-18pt)}]pic cs:upYC5) 
  node[anchor=west] {$\scriptstyle \text{$\rm D, D^c$}$}; 
\end{tikzpicture}
 &=& \frac{48}{5}\cdot\frac{15}{6}\nonumber\\ 
 &=& \boxed{24}
\eea
\bea \label{BYPUp}
(B_{YP})_{\mathrm{New}} &=& 4\bigg(\frac{3}{5}\sum_{\Phi_i}Y_{\Phi_i}\bigg)Q_{\Phi_i}^2 \nonumber\\
&=& \boxed{\frac{12}{5}\rm Tr \big[Y_{\Phi_i}^2Q_{\Phi_i}^2\big]}\\
&=& (B_{PY})_{\mathrm{New}}\nonumber
\eea
$B_{LY}, B_{LL}$ and $B_{LC}$ are similar as the MSSM calculations as shown in equations \ref{B21MSSM}, \ref{B22MSSM} and \ref{B23MSSM1} respectively.
\bea \label{BLPUp}
(B_{LP})_{\mathrm{New}} &=& 4\bigg(\frac{1}{2}\bigg)\sum_{\Phi_i}Q_{\Phi_i}^2 \big[\rm for~all~supermultiplets(\Phi_i)~that~are~doublet~under~SU(2)_L\big] \nonumber\\
&=&2\bigg[Q_{H_u}^2+Q_{H_d}^2+\tikzmark{upLP1}3\cdot\big(\tikzmark{upLP2}3Q_Q^2+Q_L^2\big)\bigg]
\begin{tikzpicture}[remember picture,overlay]
\draw[<-] 
  ([shift={(2pt,-2pt)}]pic cs:upLP1) |- ([shift={(-10pt,-12pt)}]pic cs:upLP1) 
  node[anchor=east] {$\scriptstyle \text{Generation}$}; 
  \draw[<-] 
  ([shift={(2pt,-2pt)}]pic cs:upLP2) |- ([shift={(20pt,-12pt)}]pic cs:upLP2) 
  node[anchor=west] {$\scriptstyle \text{Color}$}; 
\end{tikzpicture}
\eea
\bea \label{BCYUp}
(B_{CY})_{\mathrm{New}} &=& 4\bigg(\frac{1}{2}\bigg)\bigg(\frac{3}{5}\sum_{\Phi_i}Y_{\Phi_i}^2\bigg) \big[\rm for~all~supermultiplets(\Phi_i)~that~are~doublet~under~SU(3)_C\big] \nonumber\\
&=&\frac{6}{5}\bigg[\big(Y_Q^2+Y_u^2+Y_d^2\big)+\big(Y_D^2+Y_{D^c}^2\big)\bigg]\nonumber\\
&=&\frac{11}{5}+\big[\frac{6}{5}\cdot\tikzmark{upCY1}6\cdot\tikzmark{upCY2}1\cdot(\frac{1}{3})^2\big]\nonumber\\ 
\begin{tikzpicture}[remember picture,overlay]
\draw[<-] 
  ([shift={(2pt,-2pt)}]pic cs:upCY1) |- ([shift={(-10pt,-12pt)}]pic cs:upCY1) 
  node[anchor=east] {$\scriptstyle \text{3 pairs}$}; 
  \draw[<-] 
  ([shift={(2pt,-2pt)}]pic cs:upCY2) |- ([shift={(20pt,-12pt)}]pic cs:upCY2) 
  node[anchor=west] {$\scriptstyle \text{Singlet}$}; 
\end{tikzpicture}
&=&\frac{11}{5}+\frac{4}{5}\nonumber\\
&=&\boxed{3}
\eea
Since no new exotics transforms as doublet under SU(2), there is no extra contribution to the $B_{CL}$ element of the MSSM and thus the calculation of $(B_{CL})_{\rm New}$ is same as equation \ref{B32MSSM1}. Similar as equation \ref{B33MSSM1},
\bea \label{BCCUp}
(B_{CC})_{\mathrm{New}} &=& -6(3)^2+2\cdot 3\sum_{\Phi_i}(\frac{1}{2})\cdot\Phi_i+4\cdot\frac{1}{2}\sum_{\Phi_i}(\frac{4}{3})\cdot\Phi_i \nonumber\\ 
 &=& -54 + 2\cdot 3\cdot (\frac{1}{3}\times 18)+4\cdot\frac{1}{2}(\frac{4}{3}\times 18)\label{BCCUp1}\\ 
 &=& \boxed{48}
\eea
in equation \ref{BCCUp1}, due to the presence of exotics, $D (3)$ and $D^c (3)$, the number of supermultiplets is increased from 12 (MSSM) to 18.
\bea \label{BCPUp}
(B_{CP})_{\mathrm{New}} &=& 4\bigg(\frac{1}{2}\bigg)\sum_{\Phi_i}Q_{\Phi_i}^2 \big[\rm for~all~supermultiplets(\Phi_i)~that~are~triplet~under~SU(3)_C\big] \nonumber\\
&=&2\bigg[\tikzmark{upCP1}3\cdot\big(\tikzmark{upCP2}2Q_Q^2+Q_{u^c}^2+Q_{d^c}^2+Q_D^2+Q_{D^c}^2\big)\bigg]\nonumber\\
\begin{tikzpicture}[remember picture,overlay]
\draw[<-] 
  ([shift={(2pt,-2pt)}]pic cs:upCP1) |- ([shift={(-10pt,-12pt)}]pic cs:upCP1) 
  node[anchor=east] {$\scriptstyle \text{Generation}$}; 
  \draw[<-] 
  ([shift={(2pt,-2pt)}]pic cs:upCP2) |- ([shift={(20pt,-12pt)}]pic cs:upCP2) 
  node[anchor=west] {$\scriptstyle \text{Doublet}$}; 
\end{tikzpicture}
&=&\boxed{6\bigg[2Q_Q^2+Q_{u^c}^2+Q_{d^c}^2+Q_D^2+Q_{D^c}^2\bigg]}.
\eea
Similarly,
\bea \label{BPCUp}
(B_{PC})_{\mathrm{New}} &=& 4\bigg(\frac{4}{3}\bigg)\sum_{\Phi_i}Q_{\Phi_i}^2 \big[\rm for~all~supermultiplets(\Phi_i)~that~are~triplet~under~SU(3)_C\big] \nonumber\\
&=&\frac{16}{3}\bigg[\tikzmark{upPC1}3\cdot\tikzmark{upPC2}3\cdot\big(2Q_Q^2+Q_{u^c}^2+Q_{d^c}^2+Q_D^2+Q_{D^c}^2\big)\bigg]\nonumber\\
\begin{tikzpicture}[remember picture,overlay]
\draw[<-] 
  ([shift={(2pt,-2pt)}]pic cs:upPC1) |- ([shift={(-12pt,-14pt)}]pic cs:upPC1) 
  node[anchor=east] {$\scriptstyle \text{Generation}$}; 
  \draw[<-] 
  ([shift={(5pt,-2pt)}]pic cs:upPC2) |- ([shift={(14pt,-14pt)}]pic cs:upPC2) 
  node[anchor=west] {$\scriptstyle \text{Color}$}; 
\end{tikzpicture}
&=&\boxed{48\bigg[2Q_Q^2+Q_{u^c}^2+Q_{d^c}^2+Q_D^2+Q_{D^c}^2\bigg]}.
\eea
\bea \label{BPLUp}
(B_{PL})_{\mathrm{New}} &=& 4\bigg(\frac{3}{4}\bigg)\sum_{\Phi_i}Q_{\Phi_i}^2 \big[\rm for~all~supermultiplets(\Phi_i)~that~are~doublet~under~SU(2)_L\big] \nonumber\\
&=&3\cdot\tikzmark{upPL1}2\bigg[\tikzmark{upPL2}3\cdot\big(\tikzmark{upPL3}3Q_Q^2+Q_L^2\big)+Q_{H_u}^2+Q_{H_d}^2\bigg]\nonumber\\
\begin{tikzpicture}[remember picture,overlay]
\draw[<-] 
  ([shift={(2pt,-2pt)}]pic cs:upPL1) |- ([shift={(-10pt,-12pt)}]pic cs:upPL1) 
  node[anchor=east] {$\scriptstyle \text{Doublet}$}; 
  \draw[<-] 
  ([shift={(2pt,-2pt)}]pic cs:upPL2) |- ([shift={(20pt,-12pt)}]pic cs:upPL2) 
  node[anchor=west] {$\scriptstyle \text{Generation}$};
  \draw[<-] 
  ([shift={(5pt,-2pt)}]pic cs:upPL3) |- ([shift={(-20pt,-18pt)}]pic cs:upPL3) 
  node[anchor=east] {$\scriptstyle \text{Color}$}; 
\end{tikzpicture}
&=&\boxed{6\bigg[3\big(3Q_Q^2+Q_L^2\big)+Q_{H_u}^2+Q_{H_d}^2\bigg]}.
\eea
Finally,
\bea \label{BPPUp}
(B_{PP})_{\mathrm{New}} &=& 4\sum_{\Phi_i}Q_{\Phi_i}^4 \big[\Phi_i=\rm All~supermultiplets\big]\nonumber\\
&=&\boxed{ 4 ~\rm Tr\big[Q_{\Phi_i}^4\big]}
\eea


\section{$\beta$-function coefficients for vector-like particles (SUSY Scenario)}
\label{VLRGEs}

Now, let us consider an example when there are some vector-like particles present in the model, like in Ref. \cite{Barger:2007qb}. In that model, there exist several pairs of vector-like particles. Let's choose the particles with such quantum numbers

\begin{eqnarray}
&& XQ + {\overline{XQ}} = {\mathbf{(3, 2, {1\over 6}) + ({\bar 3}, 2,
-{1\over 6})}}\,, \quad \Delta b^Q =({1\over 5}, 3, 2)\,;\\ 
&& XU + {\overline{XU}} = {\mathbf{ ({3},
1, {2\over 3}) + ({\bar 3},  1, -{2\over 3})}}\,, \quad \Delta b^U =
({8\over 5}, 0, 1)\,;\\ 
&& XD + {\overline{XD}} = {\mathbf{ ({3},
1, -{1\over 3}) + ({\bar 3},  1, {1\over 3})}}\,, \quad \Delta b^D =
({2\over 5}, 0, 1)\,;\\  
&& XL + {\overline{XL}} = {\mathbf{(1,  2, {1\over 2}) + ({1},  2,
-{1\over 2})}}\,, \quad \Delta b^L = ({3\over 5}, 1, 0)\,;\\ 
&& XE + {\overline{XE}} = {\mathbf{({1},  1, {1}) + ({1},  1,
-{1})}}\,, \quad \Delta b^E = ({6\over 5}, 0, 0)\,;\\ 
&& XG = {\mathbf{({8}, 1, 0)}}\,, \quad \Delta b^G = (0, 0, 3)\,;\\ 
&& XW = {\mathbf{({1}, 3, 0)}}\,, \quad \Delta b^W = (0, 2, 0)\,;\\
&& XT + {\overline{XT}} = {\mathbf{(1, 3, 1) + (1, 3,
-1)}}\,, \quad \Delta b^T =({{18}\over 5}, 4, 0)\,;\\ 
&& XS + {\overline{XS}} = {\mathbf{(6,  1, -{2\over 3}) + ({\bar 6},
1, {2\over 3})}}\,, \quad \Delta b^S = ({16\over 5}, 0, 5)\,;\\ 
&& XY + {\overline{XY}} = {\mathbf{(3, 2, -{5\over 6}) + ({\bar 3}, 2,
{5\over 6})}}\,, \quad \Delta b^Y =(5, 3, 2)\,.\,
\end{eqnarray}

\subsection{1-loop $\b$ function coefficients}
In the similar way as we have calculated above, we can reproduce their results for the $\b$-function coefficients. For $\Delta b$s of vector-like fermions the contributions only come from the normalized hypercharges, 
\bea
\label{XQb1}
\Delta b_1^Q &=& \frac{3}{5}\sum_{f} Y_f^2 \nonumber\\
&=&  \frac{3}{5}\cdot\tikzmark{XQb1}3.\tikzmark{XQb2}2.\big[(\frac{1}{6})^2+(-\frac{1}{6})^2\big]\nonumber\\
\begin{tikzpicture}[remember picture,overlay]
\draw[<-] 
  ([shift={(2pt,-2pt)}]pic cs:XQb1) |- ([shift={(-14pt,-14pt)}]pic cs:XQb1) 
  node[anchor=east] {$\scriptstyle \text{Color}$};
  \draw[<-] 
  ([shift={(2pt,-2pt)}]pic cs:XQb2) |- ([shift={(14pt,-14pt)}]pic cs:XQb2) 
  node[anchor=west] {$\scriptstyle \text{Doublet}$};
\end{tikzpicture}
&=&  \boxed{\frac{1}{5}}.
\eea
Similarly,
\bea
\label{XUb1}
\Delta b_1^U &=& \frac{3}{5}\sum_{f} Y_f^2 \nonumber\\
&=&  \frac{3}{5}\cdot\tikzmark{XUb1}3\cdot\big[(\frac{2}{3})^2+(-\frac{2}{3})^2\big]\nonumber\\
\begin{tikzpicture}[remember picture,overlay]
\draw[<-] 
  ([shift={(2pt,-2pt)}]pic cs:XUb1) |- ([shift={(-14pt,-14pt)}]pic cs:XUb1) 
  node[anchor=east] {$\scriptstyle \text{Color}$};
\end{tikzpicture}
&=& \frac{3}{5}\cdot 3\cdot 2\cdot\frac{4}{9} = \boxed{\frac{8}{5}}.
\eea

\bea
\label{XLb1}
\Delta b_1^L &=& \frac{3}{5}\sum_{f} Y_f^2 \nonumber\\
&=&  \frac{3}{5}\cdot\tikzmark{XLb1}2\cdot\big[(\frac{1}{2})^2+(-\frac{1}{2})^2\big]\nonumber\\
\begin{tikzpicture}[remember picture,overlay]
\draw[<-] 
  ([shift={(2pt,-2pt)}]pic cs:XLb1) |- ([shift={(14pt,-14pt)}]pic cs:XLb1) 
  node[anchor=west] {$\scriptstyle \text{Doublet}$};
\end{tikzpicture}
&=&  \boxed{\frac{3}{5}}.
\eea

\bea
\label{XEb1}
\Delta b_1^E &=& \frac{3}{5}\sum_{f} Y_f^2 \nonumber\\
&=&  \frac{3}{5}\cdot\big[1+1\big]\nonumber\\
&=&  \boxed{\frac{6}{5}}.
\eea

\bea
\label{XTb1}
\Delta b_1^T &=& \frac{3}{5}\sum_{f} Y_f^2 \nonumber\\
&=&  \frac{3}{5}\cdot 3\cdot\big[1+1\big]\nonumber\\
&=&  \boxed{\frac{18}{5}}.
\eea

\bea
\label{XSb1}
\Delta b_1^S &=& \frac{3}{5}\sum_{f} Y_f^2 \nonumber\\
&=&  \frac{3}{5}\cdot(6\times 1)\cdot\big[(\frac{2}{3})^2+(-\frac{2}{3})^2\big]\nonumber\\
&=& \frac{3}{5}\cdot 6\cdot 2\cdot\frac{4}{9} = \boxed{\frac{16}{5}}.
\eea

\bea
\label{XYb1}
\Delta b_1^Y &=& \frac{3}{5}\sum_{f} Y_f^2 \nonumber\\
&=&  \frac{3}{5}\cdot(3\times 2)\cdot\big[(\frac{5}{6})^2+(-\frac{5}{6})^2\big]\nonumber\\
&=& \frac{3}{5}\cdot 6\cdot 2\cdot\frac{25}{36} = \boxed{5}.
\eea

For $b_2$ calculation we take help from Table \ref{tab:SU2} which lists all necessary Dynkin index $S(R)$ and Quadratic Casimir invariants $C_2(R)$ for different $SU(2)$ representations. 

\begin{table}[H]
\begin{center}
\begin{tabular}{|c|c|c|}
\hline
Representation $d(R)$, under SU(2)  & Dynkin index $S(R)$ & Quadratic Casimir invariant $C_2(R)$\\  \hline\hline
 1 & 0 & 0
\\ \hline
 2 & $\frac{1}{2}$ & $\frac{3}{4}$ \\ \hline
 3 & 2 & 2 \\ \hline
\end{tabular}
\caption{Dynkin indices and quadratic Casimir invariants under different SU(2) representations.\label{tab:SU2}}
\vspace{-0.6cm}
\end{center}
\end{table}
Clearly  the particles which are singlet under $SU(2)$ have zero contribution. Thus $\Delta b_2^U= \Delta b_2^D=\Delta b_2^E=\Delta b_2^G=\Delta b_2^S=0$. For other representations, let's calculate them explicitly.

\noindent
Now in a SUSY model, the extra fermions contribution is given by the sum of index times the multiplicity. Therefore,
\bea
\label{XQb21}
\Delta b_2^Q &=& \left(\frac{1}{2}\cdot 2\right)\cdot\tikzmark{XQb21}N_Q \\ 
&=&  \big[(3+3)\cdot S(2)\big]\nonumber\\
\begin{tikzpicture}[remember picture,overlay]
  \draw[<-] 
  ([shift={(2pt,-2pt)}]pic cs:XQb21) |- ([shift={(14pt,-12pt)}]pic cs:XQb21) 
  node[anchor=west] {$\scriptstyle \text{No. of multiplets}$};
\end{tikzpicture}
&=&  \boxed{3} ~~[S(2)=\frac{1}{2} ~\rm from~Table~ \ref{tab:SU2}].
\eea
where in Eqn.\ref{XQb21}, $\frac{1}{2}$ represents the index of the fundamental and 2 is for the vector-like pairs.
Similarly,
\bea
\label{XLb2}
\Delta b_2^L &=& N_L \nonumber\\ 
&=&  \big[(1+1)\cdot S(2)\big]=2\cdot\frac{1}{2}\nonumber\\
&=&  \boxed{1},
\eea
\bea
\label{XWb2}
\Delta b_2^W &=& N_W \nonumber\\ 
&=&  \big[1\cdot S(3)\big]=1\cdot 2\nonumber\\
&=&  \boxed{2},
\eea
\bea
\label{XTb2}
\Delta b_2^T &=& N_T \nonumber\\ 
&=&  \big[(1+1)\cdot S(3)\big]=2\cdot 2\nonumber\\
&=&  \boxed{4},
\eea
and
\bea
\label{XYb2}
\Delta b_2^Y &=& N_Y \nonumber\\ 
&=&  \big[(3+3)\cdot S(2)\big]=6\cdot\frac{1}{2}\nonumber\\
&=&  \boxed{3},
\eea

\begin{table}[H]
\begin{center}
\begin{tabular}{|c|c|c|}
\hline
Representation $d(R)$, under SU(3)  & Dynkin index $S(R)$ & Quadratic Casimir invariant $C_2(R)$\\  \hline\hline
 1 & 0 & 0
\\ \hline
 3 & $\frac{1}{2}$ & $\frac{4}{3}$ \\ \hline
 $\sbar 3$ & $\frac{1}{2}$ & $\frac{4}{3}$ \\ \hline
 6 & $\frac{5}{2}$ & $\frac{10}{3}$ \\ \hline
 $\sbar 6$ & $\frac{5}{2}$ & $\frac{10}{3}$ \\ \hline
 8 & 3 & 3 \\ \hline
\end{tabular}
\caption{Dynkin indices and quadratic Casimir invariants under different SU(3) representations.\label{tab:SU3}}
\vspace{-0.6cm}
\end{center}
\end{table}

\noindent
Again, the particles which are singlet under $SU(3)$ representation, have zero contribution. Thus $\Delta b_3^L= \Delta b_3^E=\Delta b_3^W=\Delta b_3^T=0$. For others, the contributions are the following:
\bea
\label{XQb3}
\Delta b_3^Q &=& N_Q \nonumber\\ 
&=&  \big[(2+2)\cdot S(3)\big]\nonumber\\
&=&  \boxed{2} ~~[S(3)=\frac{1}{2} ~\rm from~Table~ \ref{tab:SU3}],
\eea
\bea
\label{XUb3}
\Delta b_3^U &=& N_U \nonumber\\ 
&=&  \big[(1+1)\cdot S(3)\big]=2\cdot\frac{1}{2}\nonumber\\
&=&  \boxed{1}, 
\eea
\bea
\label{XDb3}
\Delta b_3^D &=& N_D \nonumber\\ 
&=&  \big[(1+1)\cdot S(3)\big]=2\cdot\frac{1}{2}\nonumber\\
&=&  \boxed{1}, 
\eea
\bea
\label{XGb3}
\Delta b_3^G &=& N_G \nonumber\\ 
&=&  \big[1\cdot S(8)\big]=1\cdot 3\nonumber\\
&=&  \boxed{3}, 
\eea
\bea
\label{XSb3}
\Delta b_3^S &=& N_S \nonumber\\ 
&=&  \big[(1+1)\cdot S(6)\big]=2\cdot\frac{5}{2}\nonumber\\
&=&  \boxed{5}, 
\eea
\bea
\label{XYb3}
\Delta b_3^Y &=& N_Y \nonumber\\ 
&=&  \big[(2+2)\cdot S(3)\big]=4\cdot\frac{1}{2}\nonumber\\
&=&  \boxed{2}. 
\eea

\subsection{2-loop $\beta$ function coefficients}

The two-loops $\beta$ function contributions to the SM gauge couplings from the vector-like particles in supersymmetric models are given by \cite{Barger:2007qb}
\begin{eqnarray}
\label{VectorBcoeff}
\Delta B^{XE + {\overline{XE}}}=\begin{pmatrix}
\frac{72}{25}&~
0 & ~0 \cr 0 & ~0 & ~0 \cr
0 & ~0 & ~0 \end{pmatrix}  ~,~
\Delta B^{XL + {\overline{XL}}}=\begin{pmatrix}\frac{9}{25}&
~\frac{9}{5}& ~0 \cr \frac{3}{5} & ~7 & ~0 \cr
0 & ~0 & ~0
\end{pmatrix}
\end{eqnarray}

\bea \label{Vec11}
\Delta B_{YY}^E &=& -6\cdot 0+2\cdot 0+4 \sum S(R)C(R) \nonumber\\
 &=& 4 \sum_{\Phi_i} \big (\frac{3}{5}Y_{\Phi_i}^2 \big)\big (\frac{3}{5}Y_{\Phi_i}^2 \big)~~\nonumber\\ 
 &=& \frac{36}{25} \bigg[Y_{XE}^4+Y_{\overline{XE}}^4\bigg] \nonumber\\ 
 &=& \frac{36}{25}\bigg[1+1\bigg]\nonumber\\ 
 &=& \frac{36}{25}\cdot 2\nonumber\\ 
 &=& \boxed{\frac{72}{25}}
\eea
Since both $XE$ and $\overline{XE}$ are $SU(2)$ and $SU(3)$ singlet, all the components related to those groups are zero.

\noindent
Similarly for $XL$ and $\overline{XL}$,
\bea \label{XL11}
\Delta B_{YY}^L &=& 4 \sum_{\Phi_i} \big (\frac{3}{5}Y_{\Phi_i}^2 \big)\big (\frac{3}{5}Y_{\Phi_i}^2 \big)~~\nonumber\\ 
 &=& \frac{36}{25} \bigg[Y_{XL}^4+Y_{\overline{XL}}^4\bigg] \nonumber\\ 
 &=& \frac{36}{25}\cdot\tikzmark{XL1}2\cdot\bigg[\frac{1}{16}+\frac{1}{16}\bigg]\nonumber\\
 \begin{tikzpicture}[remember picture,overlay]
\draw[<-] 
  ([shift={(2pt,-2pt)}]pic cs:XL1) |- ([shift={(14pt,-14pt)}]pic cs:XL1) 
  node[anchor=west] {$\scriptstyle \text{Doublet}$};
\end{tikzpicture}   
 &=& \frac{36}{25}\cdot\frac{1}{4}\nonumber\\ 
 &=& \boxed{\frac{9}{25}}
\eea

\bea \label{XL12}
\Delta B_{YL}^L &=& 4 \sum_{\Phi_i} \big (\frac{3}{5}Y_{\Phi_i}^2 \big)\big (\frac{3}{4}\big)~~\nonumber\\ 
 &=& \frac{9}{5} \bigg[Y_{XL}^2+Y_{\overline{XL}}^2\bigg] \nonumber\\ 
 &=& \frac{9}{5}\cdot\tikzmark{XL2}2\cdot\bigg[\frac{1}{4}+\frac{1}{4}\bigg]\nonumber\\
 \begin{tikzpicture}[remember picture,overlay]
\draw[<-] 
  ([shift={(2pt,-2pt)}]pic cs:XL2) |- ([shift={(14pt,-14pt)}]pic cs:XL2) 
  node[anchor=west] {$\scriptstyle \text{Doublet}$};
\end{tikzpicture}   
 &=& \boxed{\frac{9}{5}}
\eea

\bea \label{XL21}
\Delta B_{LY}^L &=& 4 \big(\frac{1}{2}\big)\sum_{\Phi_i}\big (\frac{3}{5}Y_{\Phi_i}^2 \big)~~\nonumber\\ 
 &=& \frac{6}{5} \bigg[Y_{XL}^2+Y_{\overline{XL}}^2\bigg] \nonumber\\ 
 &=& \frac{6}{5}\cdot 2\cdot\bigg[\frac{1}{4}+\frac{1}{4}\bigg]\nonumber\\  
 &=& \boxed{\frac{3}{5}}
\eea

\bea \label{XL22}
\Delta B_{LL}^L &=& 2\cdot 2\cdot\sum_{\Phi_i}\big (\frac{1}{2}{\Phi_i} \big)+4\cdot\frac{1}{2}\cdot\sum_{\Phi_i}\big (\frac{3}{4}{\Phi_i} \big)~~\nonumber\\ 
 &=& 4\cdot\bigg(\frac{1}{2}\cdot 2\bigg)+2\cdot\bigg(\frac{3}{4}\cdot 2\bigg) \label{XL221}\\  
 &=& \boxed{7}
\eea
where in \ref{XL221} we have used total number of multiplets as (1+1).

\noindent
Again $XL$ and $\overline{XL}$ are being color singlet, don't contribute to $SU(3)$ components and thus other elements are zero.


\section{Pati-Salam Model RGEs (non-SUSY Scenario)}
\label{PSRGEs}

In the Pati-Salam (PS) gauge symmetry \cite{Pati:1974yy}, $SU(4)_C\times SU(2)_L\times SU(2)_R$, the five matter multiplets of each SM generation along with the right-handed neutrinos can be represented in two multiplets:

\begin{table}[H]
\newlength{\mylen}
\settowidth{\mylen}{Names}
\begin{center}
\begin{tabular}{|c|c|c|c|}
\hline
Names & $SU(3)_C ,\, SU(2)_L ,\, U(1)_Y$ & $SU(4)_C ,\, SU(2)_L ,\, SU(2)_R$
\\  \hline\hline
Q & $(\>{\bf 3},\>{\bf 2}\>,\>{1\over 6})$ & \multirow{2}{*}{$(\>{\bf 4},\>{\bf 2}\>,\>{1})$}\\
 $L$&  $(\>{\bf 1},\>{\bf 2}\>,\>-1)$ &
\\  \hline
 $\sbar u$  & $(\>{\bf \overline 3},\> {\bf 1},\> -{2\over 3})$ &  \multirow{4}{*}{$(\>{\bf \overline 4},\>{\bf 1}\>,\>{2})$}
\\  $\sbar d$ & $(\>{\bf \overline 3},\> {\bf 1},\> {1\over 3})$ &  \\ 
$\sbar e$ & $(\>{\bf 1},\> {\bf 1},\>1)$ & \\ 
$\sbar \nu$ & $(\>{\bf 1},\> {\bf 1},\>0)$ & \\ \hline
\end{tabular}
\caption{SM particle multiplets under PS gauge group.\label{tab:PSparticle}}
\vspace{-0.6cm}
\end{center}
\end{table}
Clearly per family there are total 16 Weyl fermions, 4 leptons and 12 quarks (including colors). 

If we just start with the PS group, then there would be two sets of scalars needed. 
\begin{itemize}
\item One set will break down the PS group to SM group, $ SU(4)_C\times SU(2)_L\times SU(2)_R \rightarrow SU(3)_C\times SU(2)_L\times U(1)_Y $
\item The other one for the electroweak symmetry breaking, $SU(3)_C\times SU(2)_L\times U(1)_Y \rightarrow SU(3)_C\times U(1)_{em} $ 
\end{itemize} 
But, if there we consider any other higher group of which PS group is a subgroup, then another set of scalars are needed to brek that down to PS group.
The scalar sector multiplets of the PS gauge group can have different representations  in terms of their PS quantum numbers. 


\noindent
Let us consider \cite{Babu:2015bna} as our reference for this purpose. The details of the scalar sector is given in Table [1] of that reference and the relevant scalars for our calculation, at the intermediate scale $M_I$, are 
\begin{itemize}
\item A complex Higgs sector, $126_H$ that breaks the Pati-Salam symmetry to the Standard Model
\item Another complex $10_H$ scalar that does the electroweak symmetry breaking.
\end{itemize}

\begin{table}[H]
\settowidth{\mylen}{Names}
\begin{center}
\begin{tabular}{|c|c|c|}
\hline
$SU(4)_C ,\, SU(2)_L ,\, SU(2)_R$ & $SO(10)$
\\  \hline\hline
$H_D (1,\,2,\,2)$ & $10$\\
 \hline
 $\Sigma_1(6,\,1,\,1)$  &  \multirow{4}{*}{$126$}
\\  $\Sigma_2(10,\,3,\,1)$ &  \\ 
$\Sigma_3(\sbar 10,\,1,\,3)$ & \\ 
$\Sigma_4(15,\,2,\,2)$ & \\ \hline
\end{tabular}
\caption{Relevant scalar sector representations under PS gauge group.\label{tab:PSparticle}}
\vspace{-0.6cm}
\end{center}
\end{table}

\noindent
The beta-function coefficients for 1-loop and 2-loops are given as \cite{Babu:2015bna}
\be
\label{PSbcoeff}
b_{\mathrm{PS}}=\left(4C,\,2L,\,2R\right)=\left(1,\,\frac{26}{3},\,\frac{26}{3}\right), 
\ee
\be
\label{PSBcoeff}
\big(B_{ij}\big)_{\rm PS}=\begin{pmatrix}
\frac{1209}{2}&~~\frac{249}{2}& ~~\frac{249}{2} \cr  \frac{1245}{2} & ~~\frac{779}{3}& ~~48 \cr
\frac{1245}{2} & ~~48 & ~~\frac{779}{3}
\end{pmatrix}.
\ee

\subsection{1-loop $\beta$-function coefficients}

\noindent
To calculate the $b_{PS}$s we follow Eq. \ref{genbcoeff}. The Dynkin indices and the quadratic Casimir invariant of the $SU(4)$ group are collected from Ref. \cite{Yamatsu:2015npn} and the relevant ones for our purpose are listed in Table \ref{tab:SU4}.

\begin{table}[H]
\begin{center}
\begin{tabular}{|c|c|c|}
\hline
Representation $d(R)$, under SU(4)  & Dynkin index $S(R)$ & Quadratic Casimir invariant $C_2(R)$\\  \hline\hline
 1 & 0 & 0
\\ \hline
 4 & $\frac{1}{2}$ & $\frac{15}{8}$ \\ \hline
 $\sbar 4$ & $\frac{1}{2}$ & $\frac{15}{8}$ \\ \hline
 6 & 1 & $\frac{5}{2}$ \\ \hline
 10 & 3 & $\frac{9}{2}$ \\ \hline
 $\sbar {10}$ & 3 & $\frac{9}{2}$ \\ \hline
 15 & 4 & 4 \\ \hline
\end{tabular}
\caption{Dynkin indices and quadratic Casimir invariants under different SU(4) representations.\label{tab:SU4}}
\vspace{-0.6cm}
\end{center}
\end{table}

\bea
\label{b1PS}
(b_1)_{PS} &=&-\frac{11}{3}\cdot 4+\frac{4}{3}\cdot\frac{1}{\tikzmark{b1PS1}2} \bigg(\sum_{\mathrm{fermions}}S(R_f)\bigg)+\frac{1}{6}\cdot\tikzmark{b1PS2}2\cdot\bigg(\sum_{\mathrm{scalars}}S(R_s)\bigg)
\begin{tikzpicture}[remember picture,overlay]
\draw[<-] 
  ([shift={(2pt,-2pt)}]pic cs:b1PS1) |- ([shift={(-10pt,-10pt)}]pic cs:b1PS1) 
  node[anchor=east] {$\scriptstyle \text{Weyl fermion}$}; 
\draw[<-] 
  ([shift={(2pt,-2pt)}]pic cs:b1PS2) |- ([shift={(-12pt,-18pt)}]pic cs:b1PS2) 
  node[anchor=east] {$\scriptstyle \text{Complex scalar}$};   
\end{tikzpicture}
\eea 

\vspace{0.8cm}
\noindent
where $\sum_{\mathrm{fermions}}S(R_f)$ is the total number of multiplets times the Dynkin index of fermions and $\sum_{\mathrm{scalars}}S(R_s)$ is the same for scalars. We shall calculate these quantities explicitly.

\noindent
\\
Since all the SM fermions are quadruplet under $SU(4)$, they all have contribution towards counting the number of multiplets. Therefore, 
\bea
\label{b1PSf}
\sum_{\mathrm{fermions}}S(R_f)&=& 3~(\rm generation)\times  (2\cdot 1+1\cdot 2)\cdot\,S(4)\nonumber\\
&=&12\cdot\,\frac{1}{2}=6.
\eea
The scalars which are relevant for $SU(4)$ calculations are $\Sigma_1(6,\,1,\,1), \Sigma_2(10,\,3,\,1), \Sigma_3(\sbar 10,\,1,\,3)$ and $\Sigma_4(15,\,2,\,2)$. Their contributions are the following:
\bea
\label{b1PSs}
\Sigma_1(6,\,1,\,1) &=& (1\times 1)\cdot S(6) = 1\cdot 1 = 1\,,\nonumber\\
\Sigma_2(10,\,3,\,1) &=& (3\times 1)\cdot S(10) = 3\cdot 3 = 9\,,\nonumber\\
\Sigma_3(10,\,1,\,3) &=& (1\times 3)\cdot S(10) = 3\cdot 3 = 9\,,\nonumber\\
\Sigma_4(15,\,2,\,2) &=& (2\times 2)\cdot S(15) = 4\cdot 4 = 16\,\nonumber\\
\mathrm{Therefore} \sum_{\mathrm{scalars}}S(R_s)&=& 1+2\times 9+16=35.
\eea
Thus combining Eqs.\ref{b1PSf} and \ref{b1PSs} we get 
\bea
\label{b1PSfinal}
(b_1)_{PS} &=& -\frac{44}{3}+\frac{2}{3}\cdot 6+\frac{1}{3}\cdot 35\nonumber\\
&=& \boxed{1}
\eea

\noindent
Similarly, for $SU(2)_L$ case
\bea
\label{b2PSf}
\sum_{\mathrm{fermions}}S(R_f)&=& 3~(\rm generation)\times  (4.1)\cdot\,S(2)\nonumber\\
&=&12\cdot\,\frac{1}{2}=6 ~~[\rm Using~Table~\ref{tab:SU2}].
\eea
But the relevant scalars are $H_D(1,\,2,\,2),$ $\Sigma_2(10,\,3,\,1)$ and $\Sigma_4(15,\,2,\,2)$. Following Table \ref{tab:SU2} their contributions can be found as
\bea
\label{b2PSs}
H_D(1,\,2,\,2) &=& (1\times 2)\cdot S(2) = 2\cdot\frac{1}{2} = 1\,,\nonumber\\
\Sigma_2(10,\,3,\,1) &=& (10\times 1)\cdot S(3) = 10\cdot 2 = 20\,,\nonumber\\
\Sigma_4(15,\,2,\,2) &=& (15\times 2)\cdot S(2) = 30\cdot\frac{1}{2} = 15\,\nonumber\\
\mathrm{Therefore} \sum_{\mathrm{scalars}}S(R_s)&=& 1+20+15=36.
\eea
Using Eqs. \ref{b2PSf} and \ref{b2PSs} we find
\bea
\label{b2PS}
(b_2)_{PS} &=&-\frac{22}{3}+\frac{2}{3}\cdot 6+\frac{1}{3}\cdot 36\nonumber\\
&=& \boxed{\frac{26}{3}}.
\eea

\subsection{2-loops $\beta$-function coefficients}

From Eqn. \ref{genBcoeff},
\bea
\label{PS4C4C1}
(B_{11})_{PS}=B_{4C4C} &=& -\frac{34}{3}\big[C_2(G_{4C})\big]^2+\frac{1}{2}\bigg[4C_2(F_{4C})+\frac{20}{3} C_2(G_{4C})\bigg]S_2(F_{4C})\nonumber\\
&& +2\bigg[2C_2(S_{4C})+\frac{1}{3} C_2(G_{4C})\bigg]S_2(S_{4C})
\eea

\begin{table}[H]
\begin{center}
\begin{tabular}{|c|c|c|c|c|}
\hline
\multicolumn{5}{|c|}{Scalars} \\ \hline
Names &  $C_2(S_{4C})$ & $C_2(G_{4C})$ & $S_2(S_{4C})$ & No of multiplets\\  \hline\hline
$H_D(1,\,2,\,2)$ & 0 & 4 & 0 & 0
\\ \hline
$\Sigma_1(6,\,1,\,1)$ & $\frac{5}{2}$ & 4 & 1 & $1\times 1=1$
\\ \hline
$\Sigma_2(10,\,1,\,3)$ & $\frac{9}{2}$ & 4 & 3 & $1\times 3=3$
\\ \hline
$\Sigma_3(\sbar {10},\,3,\,1)$ & $\frac{9}{2}$ & 4 & 3 & $3\times 1=3$
\\ \hline
$\Sigma_4(15,\,2,\,2)$ & 4 & 4 & 4 & $2\times 2=4$
\\ \hline\hline
\multicolumn{5}{|c|}{Fermions} \\ \hline
$(4,\,2,\,1)$ & $\frac{15}{8}$ & 4 & $\frac{1}{2}$ & $2\times 1\times 3=6$
\\ \hline
$(\sbar 4,\,1,\,2)$ & $\frac{15}{8}$ & 4 & $\frac{1}{2}$ & $1\times 2\times 3=6$
\\ \hline
\end{tabular}
\caption{Values needed for calculations of SU(4) representations.\label{tab:PS4C4C}}
\vspace{-0.6cm}
\end{center}
\end{table}
Relevant fermions and scalars and their Dynkin indices and Quadratic Casimir invariants are shown in Table \ref{tab:PS4C4C}. For fermions, an extra factor of 3 is to be multiplied to take care of the generation. Therefore, using that table and from Eq. \ref{PS4C4C} we get,
\bea
\label{PS4C4C}
B_{4C4C} &=& -\frac{34}{3}[4]^2+\frac{1}{2}\big[4\cdot\frac{15}{8}+\frac{20}{3}\cdot 4\big]\frac{1}{2}\cdot(6+6)\big]+2\big[0+\big(2\cdot\frac{5}{2}+\frac{1}{3}\cdot 4\big)\cdot 1 \nonumber\\
&& +\big(2\cdot\frac{9}{2}+\frac{1}{3}\cdot 4\big)\cdot(3\cdot 3\cdot 2)+(2\cdot 4+\frac{1}{3}\cdot 4\big)\cdot 4\cdot 4 \big]\nonumber\\
&=& -\frac{544}{3}+\big[\big(\frac{15}{2}+\frac{80}{3}\big)\cdot 3\big]+2\big[(5+\frac{4}{3})+(9+\frac{4}{3})\cdot 18+(8+\frac{4}{3})\cdot 16\big]\nonumber\\
&=&  -\frac{544}{3}+\frac{205}{2}+\frac{38}{3}+372+\frac{896}{3}\nonumber\\
&=& \frac{-1088+615+76+2232+1792}{6}\nonumber\\
&=& \boxed{\frac{1209}{2}}
\eea

\bea
\label{PS2R2R1}
(B_{33})_{PS}=B_{2R2R} &=& -\frac{34}{3}\big[C_2(G_{2R})\big]^2+\frac{1}{2}\bigg[4C_2(F_{2R})+\frac{20}{3} C_2(G_{2R})\bigg]S_2(F_{2R})\nonumber\\
&& +2\bigg[2C_2(S_{2R})+\frac{1}{3} C_2(G_{2R})\bigg]S_2(S_{2R})
\eea

\begin{table}[H]
\begin{center}
\begin{tabular}{|c|c|c|c|c|}
\hline
\multicolumn{5}{|c|}{Scalars} \\ \hline
Names &  $C_2(S_{2R})$ & $C_2(G_{2R})$ & $S_2(S_{2R})$ & No of multiplets\\  \hline\hline
$H_D(1,\,2,\,2)$ & $\frac{3}{4}$ & 2 & $\frac{1}{2}$ & $1\times 2=2$
\\ \hline
$\Sigma_3(10,\,1,\,3)$ & 2 & 2 & 2 & $10\times 1=10$
\\ \hline
$\Sigma_4(15,\,2,\,2)$ & $\frac{3}{4}$ & 2 & $\frac{1}{2}$ & $15\times 2=30$
\\ \hline\hline
\multicolumn{5}{|c|}{Fermions} \\ \hline
$(\sbar 4,\,1,\,2)$ & $\frac{3}{4}$ & 2 & $\frac{1}{2}$ & $4\times 1\times 3=12$
\\ \hline
\end{tabular}
\caption{Values needed for calculations of SU(2) representations.\label{tab:PS2R2R}}
\vspace{-0.6cm}
\end{center}
\end{table}
Therefore, using Table \ref{tab:PS2R2R}, from Eq. \ref{PS2R2R1} we get,
\bea
\label{PS2R2R}
B_{2R2R} &=& -\frac{34}{3}[2]^2+\frac{1}{2}\big[4\cdot\frac{3}{4}+\frac{20}{3}\cdot 2\big]\frac{1}{2}\cdot 12+2\bigg[\big(2\cdot \frac{3}{4}+\frac{1}{3}\cdot 2\big)+\big(2\cdot 1+\frac{2}{3}\big)\cdot 2\cdot 10 \nonumber\\
&& +\big(2\cdot\frac{3}{4}+\frac{2}{3}\big)\cdot\frac{1}{2}\cdot 30\bigg]\nonumber\\
&=& -\frac{136}{3}+3\cdot\frac{49}{3}+2\cdot \big[\frac{13}{6}+\frac{14}{3}\cdot 20+\frac{13}{6}\cdot 15\big]\nonumber\\
&=& -\frac{136}{3}+49+\frac{13}{3}+\frac{560}{3}+\frac{195}{3}\nonumber\\
&=& \boxed{\frac{779}{3}}= B_{2L2L} = (B_{22})_{PS}
\eea 
For the calculation of $B_{12}=(B_{4C2L})_{PS}$ element we need the fermions and scalars which are not singlet under either $SU(4)$ or $SU(2)_L$. Therefore, the relevant fermions are with quantum numbers $(4,\,2,\,1)$ and scalars are $\Sigma_3(\sbar {10},\,3,\,1)$ and $\Sigma_4(15,\,2,\,2)$. Thus,
\bea
\label{PS4C2L1}
(B_{12})_{PS}=B_{4C2L} &=& \frac{1}{2}\bigg[4C_2(F_{2L})\bigg]S_2(F_{4C})+2\bigg[2C_2(S_{2L})\bigg]S_2(S_{4C})\nonumber\\
&=&\frac{1}{2}\big[4\cdot\frac{3}{4}\big]\frac{1}{2}\cdot \tikzmark{4C2L1}6+2\big[2C_2\big(S_{2L}(3)\big)S_2\big(S_{4C}(10)\big)\nonumber\\
&& +2C_2\big(S_{2L}(2)\big)S_2\big(S_{4C}(15)\big)\big]\nonumber\\
 \begin{tikzpicture}[remember picture,overlay]
\draw[<-] 
  ([shift={(2pt,-2pt)}]pic cs:4C2L1) |- ([shift={(14pt,-12pt)}]pic cs:4C2L1) 
  node[anchor=west] {$\scriptstyle \text{Multiplets}$};
\draw[<-] 
  ([shift={(2pt,-2pt)}]pic cs:4C2L2) |- ([shift={(-14pt,-12pt)}]pic cs:4C2L2) 
  node[anchor=east] {$\scriptstyle \text{Multiplets}$};
\draw[<-] 
  ([shift={(2pt,-2pt)}]pic cs:4C2L3) |- ([shift={(14pt,-12pt)}]pic cs:4C2L3) 
  node[anchor=west] {$\scriptstyle \text{Multiplets}$};    
\end{tikzpicture}
&=&\frac{9}{2}+2\big[2\cdot 2\cdot (3\cdot\tikzmark{4C2L2}3)+2\cdot \frac{3}{4}\cdot(4\cdot\tikzmark{4C2L3}4)\big]\nonumber\\
&=& \frac{9}{2}+120\nonumber\\
&=& \boxed{\frac{249}{2}}=(B_{13})_{PS}=B_{4C2R}.
\eea
Because, for $(B_{13})_{PS}=B_{4C2R}$, the relevant multiplets are $(\sbar 4,\,1,\,2)$, $\Sigma_2(\sbar {10},\,1,\,3)$ and $\Sigma_4(15,\,2,\,2)$ and thus they are equal to $B_{4C2L}$.

\noindent
Now, for $(B_{21})_{PS}$ the multiplets are same as that of $(B_{21})_{PS}$, but they contribute differently as shown in the following:
\bea
\label{PS2L4C1}
B_{2L4C} &=& \frac{1}{2}\bigg[4C_2(F_{4C})\bigg]S_2(F_{2L})+2\bigg[2C_2(S_{4C})\bigg]S_2(S_{2L})\nonumber\\
&=&\frac{1}{2}\big[4\cdot\frac{15}{8}\big]\frac{1}{2}\cdot\tikzmark{2L4C1}12+2\big[2C_2\big(S_{4C}(10)\big)S_2\big(S_{2L}(3)\big)\nonumber\\
&& +2C_2\big(S_{4C}(15)\big)S_2\big(S_{2L}(2)\big)\big]\nonumber\\
 \begin{tikzpicture}[remember picture,overlay]
\draw[<-] 
  ([shift={(2pt,-2pt)}]pic cs:2L4C1) |- ([shift={(14pt,-12pt)}]pic cs:2L4C1) 
  node[anchor=west] {$\scriptstyle \text{Multiplets}$};
\draw[<-] 
  ([shift={(2pt,-2pt)}]pic cs:2L4C2) |- ([shift={(-14pt,-12pt)}]pic cs:2L4C2) 
  node[anchor=east] {$\scriptstyle \text{Multiplets}$};
\draw[<-] 
  ([shift={(2pt,-2pt)}]pic cs:2L4C3) |- ([shift={(14pt,-12pt)}]pic cs:2L4C3) 
  node[anchor=west] {$\scriptstyle \text{Multiplets}$};    
\end{tikzpicture}
&=&\frac{45}{2}+2\big[2\cdot\frac{9}{2}\cdot(2\cdot\tikzmark{2L4C2}10)+2\cdot 4\cdot(\frac{1}{2}\cdot\tikzmark{2L4C3}30)\big]\nonumber\\
&=& \frac{45}{2}+600\nonumber\\
&=& \boxed{\frac{1245}{2}}=(B_{31})_{PS}=B_{2R4C}.
\eea

Finally, for the element $(B_{23})_{PS}$ we find that there is no fermion which are doublet under both $SU(2)_L$ and $SU(2)_R$. Therefore there is no contribution from the fermions and the relevant scalars are $H_D(1,\,2,\,2), \Sigma_4(15,\,2,\,2)$.
\bea
\label{PS2R2L1}
B_{2L2R} &=& 2\bigg[2C_2(S_{2L})\bigg]S_2(S_{2R})\nonumber\\
 \begin{tikzpicture}[remember picture,overlay]
\draw[<-] 
  ([shift={(2pt,-2pt)}]pic cs:2L2R1) |- ([shift={(-14pt,-12pt)}]pic cs:2L2R1) 
  node[anchor=east] {$\scriptstyle \text{Multiplets}$};
\draw[<-] 
  ([shift={(2pt,-2pt)}]pic cs:2L2R2) |- ([shift={(14pt,-12pt)}]pic cs:2L2R2) 
  node[anchor=west] {$\scriptstyle \text{Multiplets}$};   
\end{tikzpicture}
&=&2\big[2\cdot\frac{3}{4}\cdot(\frac{1}{2}\cdot\tikzmark{2L2R1}30)+2\cdot\frac{3}{4}\cdot(\frac{1}{2}\cdot\tikzmark{2L2R2}2)\big]\nonumber\\
&=& 45+3\nonumber\\
&=& \boxed{48}=(B_{32})_{PS}=B_{2R2L}.
\eea


\section{Summary}

Starting from the basic formulea we have explicitly shown the calculation for $\beta$-function coefficients which form the renormalization group equations of different models. It is also shown that these coefficients depend completely on the quantum numbers of the gauge groups they represent. The Standard Model, Minimal Supersymmetric Standard Model, an extension of Standard Model with an $\uonep$ gauge group, a model with vector-like leptons and the Pati-Salam model has been discussed. For any model it is important to know the Dynkin indices, quadratic Casimir invariants of corresponding gauge groups and the number of multiplets under those group  representations.


\section*{Acknowledgement}

The author would especially like to thank Dr. Lorenzo Calibbi for very useful discussions, comments on the draft and encouragement.

\end{document}